\documentclass[18pt]{article}
\usepackage{lineno}
\usepackage{amsmath}
\usepackage{amsfonts}
\usepackage{amssymb}
\usepackage{graphicx}
\usepackage{setspace}
\usepackage{authblk}
\usepackage{cite}
\usepackage{setspace}
\usepackage{esvect}
\usepackage[linktocpage=true]{hyperref}
\usepackage{float}
\usepackage{color}
\usepackage[dvipsnames]{xcolor}
\usepackage{hyperref}
\hypersetup{colorlinks,linkcolor={Red},citecolor={Blue},urlcolor={Blue},filecolor={Magenta}}
\usepackage{subfig}
\usepackage{caption}
\usepackage{diagbox}
\usepackage{makecell}
\usepackage{comment}
\usepackage{graphicx}
\usepackage[compat=1.1.0]{tikz-feynman}
\usepackage{tikz-feynman,contour}
\usepackage{hhline}
\begin{document}

\title{\textbf{Search for Dark Matter in association with a Higgs boson at the LHC: A model independent study} }

\author[1,2]{Sweta Baradia\footnote{swetabaradia13@gmail.com}}
\author[3]{Sanchari Bhattacharyya\footnote{sanchari1192@gmail.com}}
\author[4]{Anindya Datta\footnote{adphys@caluniv.ac.in}}
\author[1,2]{Suchandra Dutta\footnote{suchandra.dutta@cern.ch}}
\author[5]{\\Suvankar Roy Chowdhury\footnote{sroychow@cern.ch}} 
\author[1,2]{Subir Sarkar\footnote{subir.sarkar@cern.ch}}
\affil[1]{\emph{\small Saha Institute of Nuclear Physics, 1/AF Bidhan Nagar, Kolkata, 700064, India}}
\affil[2]{\emph{\small Homi Bhabha National Institute, Training School Complex, Anushaktinagar, Mumbai, 400094, India}}
\affil[3]{\emph{\small Centre for High Energy Physics, Indian Institute of Science, Bengaluru, 560012, India}}
\affil[4]{\emph{\small Department of Physics, University of Calcutta, Kolkata, 700009, India}}
\affil[5]{\emph{\small University of Petroleum and Energy Studies, Bidholi Via-Prem Nagar, Dehradun, 248007, India}}
\date{}
	
\maketitle
	
\vskip 2cm

\begin{abstract}

Astrophysical and cosmological observations strongly suggest the existence of Dark Matter\,(DM). Experiments at the Large Hadron Collider\,(LHC) have the potential to probe the particle nature of the DM. In the present work, we investigate the potential of the mono-Higgs plus Missing Transverse Energy signature at the LHC to search for a relatively light fermionic dark matter candidate using the framework of Effective Field Theory. In our study, the DM interacts with the Standard Model\,(SM) via dimension-6 and dimension-7 effective operators involving the Higgs and the gauge bosons. Although, our analysis is independent of any Ultra Violet complete dynamics of DM, such interactions can be realized in an extension of the SM where the gauge group is extended minimally by adding an extra $U(1)$.
Both cut-based and Boosted Decision Tree\,(BDT) discriminators are used to estimate and optimize the signal sensitivity over the SM backgrounds, assuming an integrated luminosity of $3000~fb^{-1}$ at $\sqrt{s}=14$ TeV at the High Luminosity phase of the LHC\,(HL-LHC). It can be seen that in the best scenario, atleast $4\sigma$ significance can be achieved for relic masses upto 200 GeV, showcasing the prospects of this search at the HL-LHC. This study provides a foundation for future explorations in this direction.
\end{abstract}
\section{Introduction}
Accumulating evidences from several independent experiments compel us to think about the existence of a new form of matter, namely the Dark Matter\,(DM), which cannot be accommodated in the paradigm of the Standard Model\,(SM). Experimental data from the PLANCK satellite~\cite{Planck} conclusively reveal that DM constitutes about 25\% of the energy density of the Universe. The remarkably large fraction of dark matter, compared to luminous matter, is also supported by the analysis of CMBR anisotropy from WMAP~\cite{WMAP}. Satellite based experiments like PAMELA~\cite{PAMELA}, FermiLAT~\cite{Fermilat} and AMS~\cite{AMS} are also looking for indirect evidences in support of the presence of a non-luminous matter in the Universe.
However, the exact nature of the DM is still an enigma, apart from the fact that it interacts with the normal matter via gravity.

Unveiling the nature of DM remains a challenge for the particle physics community, as the SM lacks a viable candidate despite strong belief in the particle nature of the DM. Initially, neutrinos were thought to play the role of the relic particle. However, relativistic nature of the neutrinos makes them unfit for structure formation of the Universe. Consequently, one has to look beyond the SM\,(BSM) for possible solution  of the DM problem. Particle physics models with Supersymmetry\,(SUSY)~\cite{SUSY-DM1, SUSY-DM2} and extra dimensions~\cite{extrad} have been theoretically very attractive as both of these large class of models can provide a viable DM candidate apart from offering remedies to many other shortcomings of the SM. Apart from SUSY, Little Higgs model~\cite{little-higgs}, Left-Right Symmetric Models~\cite{LRDM1, LRDM2}, models with extended scalar sector~\cite{extended-scalar}, or an $U(1)$ extended SM~\cite{extrau1} have also been under scrutiny of both the theorists and experimentalists. These theoretical models have been extensively studied and their possible signatures have been looked for at the LHC. Unfortunately, non-observation of any signature at the LHC pertaining to these models only pushes the limits on the masses of the DM candidates to TeV and higher~\cite{LHC-limits}. One of the possible implications of this results is that any possible dynamics of DM is beyond the reach of the LHC\footnote{In the alternative scenario, DM is coupled to the SM sector only  weakly and we have to wait for more and more data which will be available to us in the HL-LHC run in the future.} and it affects the experimental data only through some distortion of kinematical distribution of minimally modifying the rates of the processes involving SM particles.

A few conclusions can be drawn from the above observations. The DM may have an origin entirely decoupled from the SM, interacting only through gravity. Alternatively, a more comprehensive framework could include the SM as a subset, with additional heavy degrees of freedom beyond the reach of LHC. In this scenario, the DM is part of the same framework, with its interactions with the SM mediated by one or more heavy fields.
Like many particle physics models of DM~\cite{DM-models1, DM-models2, DM-models3}, we will also assume that the relic particle is void of any SM charge, and that it interacts with the SM sector via heavy BSM fields. 
Without going into the details of an Ultra-Violet\,(UV) complete model, the most straightforward way to describe such interactions is through effective operators of dimension higher than four. These operators describe interactions between the relic particle and SM particles while ensuring that all SM symmetries are maintained. Such an approach is not very uncommon. Authors in~\cite{DM_eff1, DM_eff2} have reviewed such possible operators for several variants of the relic particle. The particle sector has been extended  minimally paving the way for several variants of interactions among relic and the SM particles. In the present article, we are particularly interested in a scenario with a fermionic relic that interacts with the SM via higher dimensional operators involving the Higgs boson. This approach is the same as the Effective Field Theory\,(EFT) way to parameterize New Physics\,(NP). Several authors have used this approach to propose and study possible signatures of DM. For example, in~\cite{monoh1, monoh2, monoh3}, dimension-6\,(dim-6) and dimension-7\,(dim-7) operators were constructed out of SM fields and a Dirac fermion posing as the relic. Authors in~\cite{monoh1} have studied mono-Higgs plus Missing Transverse Energy\,(MET) final state in pp collision at the LHC as a probe of relic particle production. In a more recent analysis~\cite{monoh4, monoX_2DM}, higher dimensional operators involving relic\,(scalar and fermion) with leptons and gauge bosons have been considered. Final states comprising a photon or $Z$-boson have been considered in the future $e^+ e^-$ collider. However, these analyses have not considered how the unknown coefficients of the operators fare with the measured relic density or the experimental upper limit on the direct detection cross-section for relic particle-nucleon scattering.

Previously, many authors~\cite{extrau1, fermion-DM1, fermion-DM2, fermion-DM3} have studied the prospect of a DM with spin-$\frac{1}{2}$, both Dirac and Majorana in nature, with their masses varying from sub-MeV scale to TeV scale. All such studies involve a plethora of other BSM fields. Looking for a DM signal in a model dependent framework also requires a careful consideration of the limits on masses and couplings of such heavy fields with the SM particles.
For such a scenario, a suitable framework for exploring BSM physics is EFT \cite{dark_eft1, dark_eft2, DM_eff1, DM_eff2}.
It is important to ensure that the effective coupling\,(s) between the relic and the SM sector result in a relic density and direct detection cross-section in the right ballpark.

So, our aim is two-fold. First, we seek to determine the range of effective coupling values that correspond to the correct relic density while also satisfying the direct detection cross-section constraints for DM interactions with standard matter. Next, using these effective coupling values, we investigate the potential signatures of DM arising from such interactions at the LHC.

Before delving into the details of the framework and the analysis, let us briefly highlight the main features of our analysis.

\begin{itemize}
\item We have considered the effect of relic density measurements as well as the upper limit on the cross-section of direct detection on the Wilson coefficients of the dim-6 and dim-7 operators that are defined below. 

\item The allowed values\,(from relic density and direct detection measurement) of the Wilson coefficients have been used to estimate the signal strength of mono-Higgs plus MET final state at the LHC.

\item The effect of pileup\,(PU) is incorporated for all the  processes in order to simulate the overlap of multiple $pp$ interactions in the same event, as expected in the HL-LHC conditions. This makes the analysis more realistic.

\item In the collider analysis, a cut-based as well as machine learning based algorithm have been used to estimate the signal significance.

\end{itemize}

The paper is organized as the following. Section\,~\ref{sec2} describes the framework which forms the foundation of the present analysis, followed by a discussion on the constraints arising from the DM aspects. In Section\,~\ref{sec3}, the signal and the SM background processes are described. \mbox{Section\,~\ref{sec3.1}} consists of the analysis strategy followed by cut-based and machine learning based analysis in Section\,~\ref{sec3.2}. Finally, we summarize our findings in Section\,~\ref{sec4}.

\section{Description of the Framework and Related Dark Matter Aspects} \label{sec2}
The present framework consists of the SM augmented only by a Dirac fermion, $\chi$, which plays the role of the relic particle. Such a particle spectrum at the TeV scale possibly evolves from a UV complete framework with a larger symmetry group and additional particles non-trivially transforming under such symmetry. Such assumption modifies the SM Lagrangian with new effective interactions which are suppressed by heavy mass scale where we believe the NP lies. We will now focus on how these higher dimensional operators may explain the experimental results for relic density and direct detection cross-sections of the relic-nucleon scattering.

The DM candidate, which is a SM singlet, does not interact directly with SM particles. We add new interaction terms for DM candidate and SM particles in the form of dim-6 and dim-7 effective operators. The effective interactions involving $\chi$, SM Higgs doublet field $\Phi$, $SU(2)_L$ and $U(1)_Y$ gauge field strengths $W^{\mu \nu}_a$, $B^{\mu \nu}$ and SM fermions $f$, are given below,
\begin{eqnarray}
    \label{efftop1}
	\mathcal{O}_1 &=& \dfrac{1}{\Lambda^2} \left( \bar{\chi} \gamma_\mu\,( C_1^V + C_1^A\gamma^5) \chi \right) \left(\Phi^\dagger \overleftrightarrow{D^\mu} \Phi \right) \\
    \label{efftop2}
    \mathcal{O}_2 &=& \dfrac{1}{\Lambda^2} \left( C_2^V  \,\bar{\chi} \gamma_\mu\,\chi \; \bar{f} \gamma_\mu\, f + C_2^A  \,\bar{\chi} \gamma_\mu \gamma_5 \,\chi  \; \bar{f} \gamma_\mu \gamma_5\, f \right) \\
    \label{efftop3}
	\mathcal{O}_3 &=& \dfrac{C_3}{\Lambda^3} \left( \bar{\chi} \sigma_{\mu \nu} \chi \right) B^{\mu \nu} ~\left(\Phi^\dagger \Phi \right) \\
    \label{efftop4}
    \mathcal{O}_4 &=& \dfrac{C_4}{\Lambda^3} \left(\bar{\chi} \sigma_{\mu \nu} \chi \right) W^{\mu \nu}_a ~\left(\Phi^\dagger \tau_a \Phi \right) 
\end{eqnarray}
where, $\Lambda$ is the scale of NP. One may also add a couple of additional operators where the vector and axial vector currents in ${\cal O}_1$ and ${\cal O}_2$ will be replaced by scalar and pseudo-scalar currents. But with a minimalist viewpoint, we shall restrict ourselves to the aforementioned set of operators.

Before we conclude this section, let us discuss in a qualitative manner the possible origin of the effective operators used in our analysis. A possible UV completion of the operators used in our analysis could be realized in a simple gauged $U(1)$ extension of the SM augmented by a Dirac fermion which is void of any SM charge. The Dirac fermion identified with $\chi$ acts as the relic particle. As $\chi$ is a singlet under SM gauge group, it cannot interact with the SM particle directly. However, $\chi$ can transform non-trivially under the extra $U(1)'$, thus interacting with the extra gauge boson~\cite{dark_photon1, dark_photon2, dark_photon3, zprime_dm1}. A possible $U(1)$ mixing between the SM $U(1)_Y$ and $U(1)'$ can act as a portal through which $\chi$ can interact with the SM sector. Non-observation of any such extra gauge bosons has set the lower limit on additional gauge boson mass to 1 TeV or higher \cite{dark_photon2, zprime_dm1}. Thus, integrating out such a heavy field from the Lagrangian would result in above set of operators. In Fig.~\ref{feynman_fulltheory} are listed the Feynman diagrams of the processes which in turn lead to the effective interactions described by higher dimensional operators, where the scale $\Lambda$ is connected to the breaking of the $U(1)'$. Scalar interactions among relic particle, $\chi$, and SM sector, may also arise in such an UV complete theory, 
but in this analysis we focus on the class of operators consisting of vector and tensor interaction. 

\begin{figure}[H]
	\centering

    \subfloat[]{
    \begin{minipage}{0.55\textwidth}
	\begin{tikzpicture}
		\begin{feynman}
			\vertex(a) at (-2,-2) {\(\chi\)};
			\vertex(b) at (0,0);
			\vertex(c) at (-2,2) {\(\chi\)};
			\vertex[dot](d) at (2,0) {\contour{white}{}};
			\vertex(e) at (4,0);
			\vertex(f) at (6,0) {\(B,W^3\)};
			\vertex(f1) at (6,2) {\(\Phi\)};
			\vertex(f2) at (6,-2) {\(\Phi\)};
			\diagram* {(a) -- [fermion] (b), (b) -- [anti fermion] (c),(b) -- [boson,edge label'=\(B'\)] (d),(d) -- [boson,edge label'=\(B\)] (e),(e) -- [boson] (f),(e) -- [scalar] (f2),(e) -- [scalar] (f1)};
		\end{feynman}
	\end{tikzpicture}
	\end{minipage} 
	\begin{minipage}{0.1\textwidth}
		\begin{tikzpicture}
		\begin{feynman}
		\vertex(a) at (-1.5,0);
		\vertex(b) at (0,0);
		\vertex(c) at (-0.8,0.5);
		\vertex(d) at (-0.8,-0.5);
		\diagram* {(a) -- (b), (b) -- (c),(b) -- (d)};
		\end{feynman}
		\end{tikzpicture}
	\end{minipage}
    \begin{minipage}{0.3\textwidth}
        \begin{tikzpicture}
        \begin{feynman}
        \vertex(a) at (-2,-2) {\(\chi\)};
        \vertex[blob](b) at (0,0) {\contour{white}{}};
        \vertex(c) at (-2,2) {\(\chi\)};
        \vertex(e) at (2,0) {\(B,W^3\)};
        \vertex(f1) at (2,2) {\(\Phi\)};
        \vertex(f2) at (2,-2) {\(\Phi\)};
        \diagram* {(a) -- [fermion] (b), (b) -- [anti fermion] (c),(b) -- [boson] (e),(b) -- [scalar] (f2),(b) -- [scalar] (f1)};
        \end{feynman}
        \end{tikzpicture}
    \end{minipage}
    }
    \phantomcaption
    \end{figure}

    \begin{figure}[H]
	\centering
        \ContinuedFloat
        \subfloat[]{
        \begin{minipage}{0.55\textwidth}
			\begin{tikzpicture}
			\begin{feynman}
			\vertex(a) at (-2,-2) {\(\chi\)};
			\vertex(b) at (0,0);
			\vertex(c) at (-2,2) {\(\chi\)};
			\vertex[dot](d) at (2,0) {\contour{white}{}};
			\vertex(e) at (4,0);
			\vertex(f1) at (6,2) {\(q\)};
			\vertex(f2) at (6,-2) {\(q\)};
			\diagram* {(a) -- [fermion] (b), (b) -- [anti fermion] (c),(b) -- [boson,edge label'=\(B'\)] (d),(d) -- [boson,edge label'=\(B\)] (e),(e) -- [fermion] (f2),(e) -- [fermion] (f1)};
			\end{feynman}
			\end{tikzpicture}
        \end{minipage}
        \begin{minipage}{0.1\textwidth}
			\begin{tikzpicture}
			\begin{feynman}
			\vertex(a) at (-1.5,0);
			\vertex(b) at (0,0);
			\vertex(c) at (-0.8,0.5);
			\vertex(d) at (-0.8,-0.5);
			\diagram* {(a) -- (b), (b) -- (c),(b) -- (d)};
			\end{feynman}
			\end{tikzpicture}
        \end{minipage}
        \begin{minipage}{0.3\textwidth}
			\begin{tikzpicture}
			\begin{feynman}
			\vertex(a) at (-2,-2) {\(\chi\)};
			\vertex[blob](b) at (0,0) {\contour{white}{}};
			\vertex(c) at (-2,2) {\(\chi\)};
			\vertex(f1) at (2,2) {\(q\)};
			\vertex(f2) at (2,-2) {\(q\)};
			\diagram* {(a) -- [fermion] (b), (b) -- [anti fermion] (c),(b) -- [fermion] (f2),(b) -- [fermion] (f1)};
			\end{feynman}
			\end{tikzpicture}
        \end{minipage}
        }
        \phantomcaption
    \end{figure}

    \begin{figure}[H]
	\centering
        \ContinuedFloat
        \subfloat[]{
		\begin{minipage}{0.55\textwidth}
		\begin{tikzpicture}
		\begin{feynman}
		\vertex(a) at (-2,-2) {\(\chi\)};
		\vertex(b) at (0,0);
		\vertex(c) at (-2,2) {\(\chi\)};
		\vertex(l1) at (-1,-1);
		\vertex(l2) at (-1,1);
		\vertex[dot](d) at (2,0) {\contour{white}{}};
		\vertex(e) at (4,0);
		\vertex(f) at (6,0) {\(B,W^3\)};
		\vertex(f1) at (6,2) {\(\Phi\)};
		\vertex(f2) at (6,-2) {\(\Phi\)};
		\diagram* {(a) -- [fermion] (b), (b) -- [anti fermion] (c),(b) -- [boson,edge label'=\(B'\)] (d),(d) -- [boson,edge label'=\(B\)] (e),(e) -- [boson] (f),(e) -- [scalar] (f2),(e) -- [scalar] (f1),(l1) -- [edge label'=\(B'\), boson] (l2)};
		\end{feynman}
		\end{tikzpicture}
	\end{minipage}
	\begin{minipage}{0.1\textwidth}
		\begin{tikzpicture}
		\begin{feynman}
		\vertex(a) at (-1.5,0);
		\vertex(b) at (0,0);
		\vertex(c) at (-0.8,0.5);
		\vertex(d) at (-0.8,-0.5);
		\diagram* {(a) -- (b), (b) -- (c),(b) -- (d)};
		\end{feynman}
		\end{tikzpicture}
	\end{minipage}
	\begin{minipage}{0.3\textwidth}
		\begin{tikzpicture}
		\begin{feynman}
		\vertex(a) at (-2,-2) {\(\chi\)};
		\vertex[blob](b) at (0,0) {\contour{white}{}};
		\vertex(c) at (-2,2) {\(\chi\)};
		\vertex(e) at (2,0) {\(B,W^3\)};
		\vertex(f1) at (2,2) {\(\Phi\)};
		\vertex(f2) at (2,-2) {\(\Phi\)};
		\diagram* {(a) -- [fermion] (b), (b) -- [anti fermion] (c),(b) -- [boson] (e),(b) -- [scalar] (f2),(b) -- [scalar] (f1)};
		\end{feynman}
		\end{tikzpicture}
	\end{minipage}
	}	
	\caption{Feynman diagrams representing the interactions from a possible full theory e.g., $U(1)'$ extended gauge model, and the corresponding effective vertices at a low energy compared to the mass scale of the NP. Top (a), middle (b) and bottom (c) diagrams describe the possible origin of interactions for $\mathcal{O}_1$, $\mathcal{O}_2$ and $\mathcal{O}_3/\mathcal{O}_4$ respectively. The `dot' denotes the mixing between NP\,(additional $U(1)'$ in an UV complete model) and SM\,($U(1)_Y$ of the SM) whereas the `blob' stands for an effective vertex.}
	\label{feynman_fulltheory}
\end{figure}


It is evident from the discussion in the previous paragraph that the operators $\mathcal{O}_3$ and $\mathcal{O}_4$ evolve via the same mechanism in an UV complete theory. For all practical purposes, their effective couplings will be in the same ballpark. Consequently, their  Wilson coefficients\, $C_3$ and  $ C_4$, can be chosen to be of the same order \footnote{In fact, in the UV model, we have discussed, the ratio of the Wilson coefficients of these two operators can be approximately $\dfrac{g'}{g}$ which is close to $0.57 \,(\tan \theta_W)$}. Therefore, throughout our study, $\mathcal{O}_3$ and ${\cal O}_4$ will be  considered together. Consequently, a pair of $\chi$ can interact with the SM particles through a $Z$ or photon exchange proportional to $\dfrac{v^2}{\Lambda ^3} \left( C_3 \sin\theta_W + C_4\cos\theta_W \right)$ and $\dfrac{v^2}{\Lambda ^3} \left( C_3 \cos\theta_W - C_4\sin\theta_W \right)$, respectively.

Before going into a detailed study of the production and detection of the DM candidates, let us discuss some DM aspects of this theory in the next section.	

\subsection{Parameter Constraints from Dark Matter Experiments}
The measurement by PLANCK collaboration restricts the density of invisible matter discussed earlier, $\Omega h^2$, in a band with a central value of 0.12 and a spread of $\pm0.001$~\cite{Planck}. This measurement has far reaching consequences not only on several cosmological scenarios but also on the BSM frameworks. Several theoretical proposals have been made to explain the measured relic density assuming the particle nature of the DM, and many of these ideas have also been tested at the LHC. In the present analysis, a massive Dirac-fermions, $\chi$, is considered as the DM.

There are three experimental avenues for DM searches. In the direct detection experiments, signatures for scattering of DM with normal matter is looked for. In collider experiments, one searches for the signature of DM production from normal matter. Direct detection experiments like XENON~\cite{xenon1, xenon2, xenon3, xenon4}, PICO~\cite{pico}, LUX~\cite{lux}, LZ \cite{lz1, lz2} have set an upper limit on DM-nucleon scattering cross-section from the non-observation of any such scattering events. On the other hand, the PLANCK satellite based experiment has measured the relic abundance of our Universe at the present epoch analyzing the measured anisotropy of the CMBR spectra. For a particle physics framework of the DM, the model parameters must satisfy the aforementioned constraints. In the following section, we shall present and compare the predictions for direct detection cross-section and relic density in our framework with the available experimental data which enable us to have an allowed range of values of the unknown coefficients of the higher dimensional operators defined in Eqns.~\ref{efftop1}-\ref{efftop4}. It should be mentioned that satellite based experiments like PAMELA~\cite{PAMELA}, FermiLAT~\cite{Fermilat} and AMS~\cite{AMS} constitute the third category of experiments, aiming to detect the signal from annihilation of DM particles into SM particles. While such experimental data are available, we do not consider them in our analysis.

\subsubsection{Direct Detection and Relic Density}	
The relic particle $\chi$ can scatter off nucleons\,(see Fig.~\ref{feynman_DD}), mediated by a $Z$-boson\,(also $\gamma$ in case of $\mathcal{O}_3$ and $\mathcal{O}_4$) and driven by the operators ${\cal O}_1$, ${\cal O}_3$ and ${\cal O}_4$ along with the four-fermi interaction described by ${\cal O}_2$. We will consider the effect of the operators ${\cal O}_1$ and ${\cal O}_2 $ together as they are of the same order of $\Lambda^{-2}$ while effects of ${\cal O}_3$ and ${\cal O}_4$ will be considered together as they lead to similar kind of interactions with strength proportional to $\Lambda^{-3}$. 
Although, $\chi - N$ elastic scattering amplitude mediated by ${\cal O}_1$ includes an additional $Z$-boson propagator compared to the amplitude from ${\cal O}_2$, it also carries an extra factor of $v^2$\,($v$ is the vev of the Higgs boson). This factor partially compensates for the propagator suppression, making it comparable to the four-fermi amplitude. Scattering of $\chi$ with the nucleons can be both Spin-Independent\,(SI) and Spin-Dependent\,(SD). The expressions for the SI and SD scattering cross-sections, mediated by the aforementioned interactions in the extreme non-relativistic limit\,($q^2 \ll m_Z^2$, where $q$ is the momentum transfer in the scattering)~\cite{DDref}, can be written as follows:

\begin{eqnarray}
\label{si}
    \left(\sigma_{SI}\right)_V &\simeq& \dfrac{\mu_{\chi N}^2}{\pi} \left[\tilde{f}_p Z + \tilde{f}_n\,(A-Z)\right]^2  \\
    \left(\sigma_{SD}\right)_A &\simeq& \dfrac{4 \mu_{\chi N}^2}{\pi} J_N\,(J_N+1) \left[\dfrac{\langle S_p \rangle}{J_N} \tilde{a}_p + \dfrac{\langle S_n \rangle}{J_N}\tilde{a}_n \right]^2
    \label{sd}
\end{eqnarray}

For operators ${\cal O}_1$ and ${\cal O}_2$ , 
\begin{eqnarray}
    \tilde{f}_{p} &=& \dfrac{1}{\Lambda^2}\; \left( \dfrac{C^V_1 \cos \theta_W}{2g} ( 1 - 4\sin^2 \theta_W) +  C_2^V \right)  \nonumber \\
    \tilde{f}_{n} &=& \dfrac{1}{\Lambda^2}\; \left( - \dfrac{C^V_1 \cos \theta_W}{2g} + C_2^V \right)  \nonumber \\
    \tilde{a}_{p,n} &=& \dfrac{1}{\Lambda^2}\left[ \; \left( \dfrac{C^A_1 \cos \theta_W}{8g} +  C_2^A \right) \; \left(\Delta d ^{p,n} +  \Delta s^{p,n} \right) +  \left( - \dfrac{C^A_1 \cos \theta_W}{8g} +  C_2^A \right) \Delta u^{p,n} \right] \nonumber
\end{eqnarray},

while for operators ${\cal O}_3$ and ${\cal O}_4$,
\begin{eqnarray}
    \tilde{f}_{p} &=& \dfrac{1}{\Lambda^3}\; \left[ \dfrac{\cos \theta_W}{2g} ( 1 - 4\sin^2 \theta_W) \, \left(C^3 \sin\theta_W + C^4 \cos\theta_W \right) + g \sin\theta_W \;\left(C^3 \cos\theta_W - C^4 \sin\theta_W \right) \right] \nonumber \\
    \tilde{f}_{n} &=& - \dfrac{1}{\Lambda^3}\; \dfrac{\cos \theta_W}{2g} \, \left(C^3 \sin\theta_W + C^4 \cos\theta_W \right)  \nonumber \\
    \tilde{a}_{p,n} &=& \dfrac{1}{\Lambda^3}\left[ \; \left( \dfrac{C^A_1 \cos \theta_W}{8g} +  C_2^A \right) \; \left(\Delta d ^{p,n} +  \Delta s^{p,n} \right) +  \left( - \dfrac{C^A_1 \cos \theta_W}{8g} +  C_2^A \right) \Delta u^{p,n} \right]  \nonumber
\end{eqnarray}
    
$\mu_{\chi N}$ denotes the reduced mass of the DM-nucleon system. $J_N$ and $\langle S_{p,n} \rangle$ stand for the spin of the nucleus and the average spin of the nucleons respectively~\cite{DDref}. $\Delta u, ~\Delta d, ~\Delta s$ are the quark-distribution functions and can be obtained, for example, from~\cite{DDref}. For an estimation of the scattering cross-section, the relevant interactions have been first implemented in \texttt{FeynRules}~\cite{feynrules} and then \texttt{micrOMEGAs6.0.5}~\cite{micromegas} is used to study the DM aspects of $\chi$.

\begin{figure}[H]
	\begin{center}
	\begin{tikzpicture}
		\begin{feynman}
			\vertex(a) {\(\chi\)};
			\vertex[below right=2 cm of a] (b);
			\vertex[blob] (b) at (1.414,-1.414) {\contour{white}{}};
			\vertex[above right=2 cm of b] (c){\(\chi\)};
			\vertex[below=2 cm of b] (d);
			\vertex[below right=2 cm of d] (e){\(q\)};
			\vertex[below left=2 cm of d] (f){\(q\)};
			\diagram* {(a) -- [fermion] (b) -- [fermion] (c),(b) -- [boson,edge label'=\(Z/\gamma\)] (d),(d) -- [fermion] (e),(d) -- [anti fermion] (f)};
		\end{feynman}
	\end{tikzpicture}
    \begin{tikzpicture}
			\begin{feynman}
			\vertex(a) at (-2.5,-2.5) {\(q\)};
			\vertex[blob](b) at (0,0) {\contour{white}{}};
			\vertex(c) at (-2.5,2.5) {\(\chi\)};
			\vertex(f1) at (2.5,2.5) {\(\chi\)};
			\vertex(f2) at (2.5,-2.5) {\(q\)};
			\diagram* {(a) -- [fermion] (b), (b) -- [anti fermion] (c),(b) -- [fermion] (f2),(b) -- [fermion] (f1)};
			\end{feynman}
			\end{tikzpicture}
	\end{center}
	\caption{Feynman diagrams representing the interactions responsible for direct detection of $\chi$ via $Z$ boson or photon mediator\,(left) and four-Fermi Effective operator\,(right).}
	\label{feynman_DD}
\end{figure}

We are now equipped with necessary tools to discuss the results of scattering of the DM off the nucleus\,(nucleons). Let us start with the SI scattering driven via ${\cal O}_1$ and ${\cal O}_2$. A careful look into the scattering rate\,(see Eqn.\ref{si}) reveals that the contributions of $C_1^V$ and $C_2^V$ in DM-neutron SI scattering cross-section\,($\sigma^n _{SI}$) have opposite sign whereas in DM-proton SI scattering cross-section\,($\sigma^p _{SI}$), they add up. With moderate values of such couplings such that these interactions can be probed at the LHC, it is very challenging to satisfy the limits coming from direct detection experiments for both protons and neutrons separately at the same time. In order to alleviate this problem and to satisfy the constraint on $\sigma^p _{SI}$ and $\sigma^n _{SI}$ simultaneously, $C_1^V$ and $C_2^V$ are both chosen to be zero in this analysis.

\begin{figure}[H]
	\begin{center}
		\includegraphics[width=8.6cm, height=6.8cm]{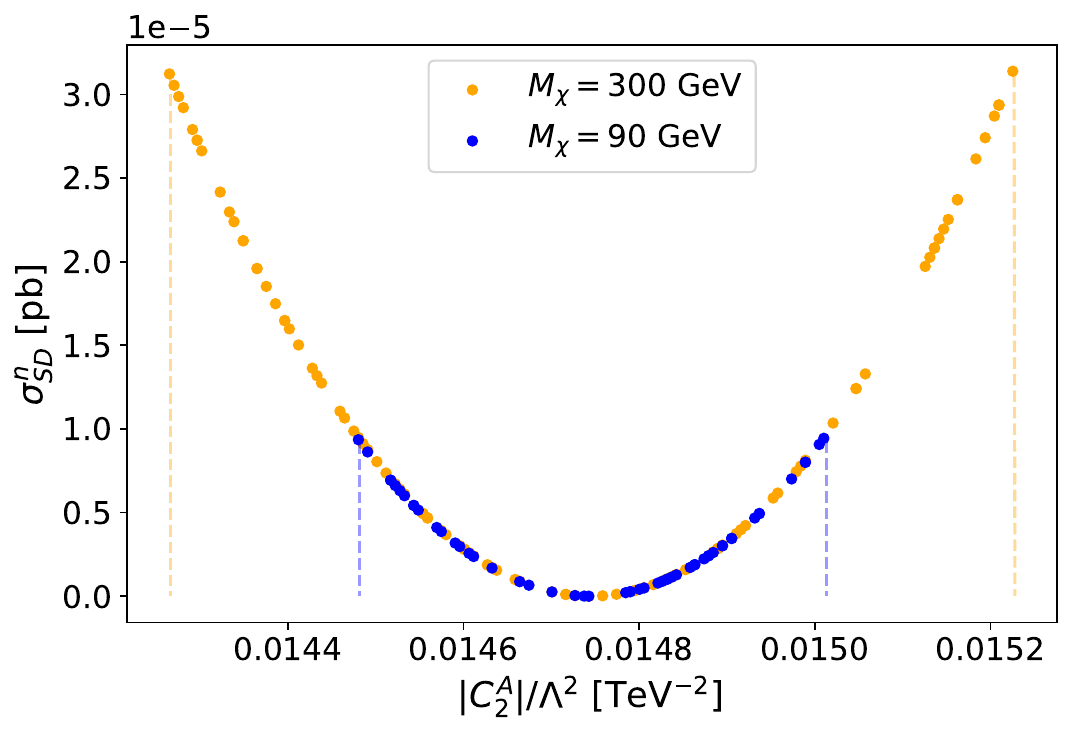}
        \includegraphics[width=8.6cm, height=6.6cm]{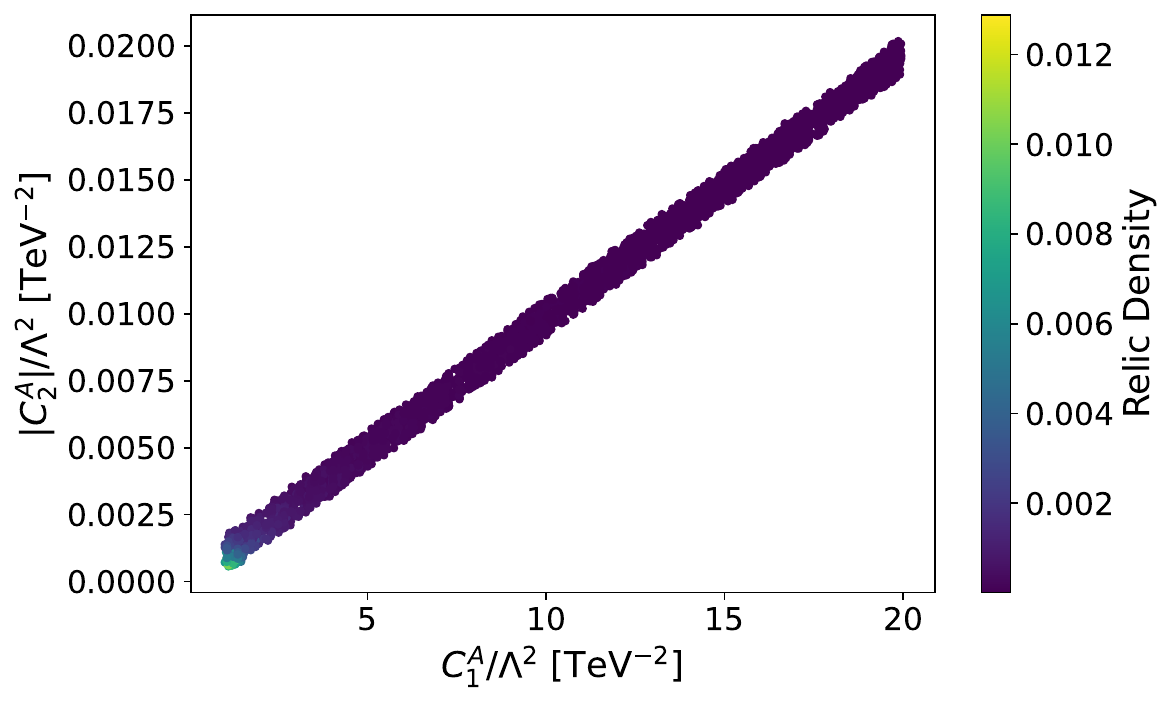}
	 \end{center}
 \caption{The left plot shows the variation of DM-neutron SD scattering cross-section\,($\sigma_{SD}^n$) with $|C_2^A|/\Lambda^2$ for ${\cal O}_1$ and ${\cal O}_2$ with $C_1^A/\Lambda^2 = 15$ TeV$^{-2}$, for two values of $M_\chi$. The points shown are within the XENONnT experiment's limit on $\sigma_{SD}^n$ \cite{xenon1}. The dashed lines show the allowed ranges of $|C_2^A|/\Lambda^2$. The plot on the right illustrates the allowed parameter space in the $C_1^A/\Lambda^2$ - $|C_2^A|/\Lambda^2$ plane, with the color bar representing the corresponding relic density values.}
 \label{dir_detect_o1}
\end{figure}

On the other hand, the SD cross section driven by ${\cal O}_1$ and ${\cal O}_2$ satisfies the experimental limits from XENON and PICO. The variation of DM-nucleon SD scattering cross-section with $|C_2^A|/\Lambda^2$  has been presented for a fixed value of $C_1^A/\Lambda^2= 15 ~\rm TeV^{-2}$ in the left panel of Fig.~\ref{dir_detect_o1} for a couple of values of $M_\chi$. The blue\,(orange) dots represent the case for $M_{\chi}=90 ~(300)$ GeV. The range of the values of $|C_2^A|/\Lambda^2$ shown in the plot is allowed by the upper limit on DM-neutron scattering cross-section by XENONnT~\cite{xenon1}. The higher the mass of DM, more relaxed is the range of $|C_2^A|/\Lambda^2$.  Projected limits from XENONnT 20 T-yr \cite{xenon3} data or limits from LZ experiment \cite{lz1, lz2} on direct detection cross-sections will constrain the allowed range of $|C_2^A|/\Lambda^2$ more stringently than the present work. However, considering all such experimental data, a value of $|C_2^A|/\Lambda^2$ around $0.0147$ TeV$^{-2}$ is allowed for all the DM masses considered as Benchmark Points\,(BPs) for the collider analysis, discussed in the next section. This motivates us to fix $|C_2^A|/\Lambda^2$ at $1.474 \times 10^{-2}$ TeV$^{-2}$ for probing the interactions arising from $\mathcal{O}_1$ at the LHC with $C_1^A/\Lambda^2 = 15$ TeV$^{-2}$.  By similar consideration, for $C_1^A/\Lambda^2 = 8.75$ TeV$^{-2}$, the value of $|C_2^A|/\Lambda^2 = 8.6 \times 10^{-3}$ TeV$^{-2}$ has been chosen for all DM masses considered as BPs. In the right panel of Fig.~\ref{dir_detect_o1}, we have presented the XENONnT allowed values of $C_1^A/\Lambda^2$ and $|C_2^A|/\Lambda^2$ with colored dots representing different values for $\Omega h^2$ for a whole range of relic particle masses in the range of 10-800 GeV. Although, the parameter points presented in the aforementioned figure satisfy direct detection constraint, their contribution to relic density are very small. Thus, such a BSM framework cannot explain the relic density of the Universe on its own. 

We also obtain results for the direct detection cross-section driven by the operators $\mathcal{O}_3$ and $\mathcal{O}_4$. The following discussion presents two cases: $C_3 = C_4$ and $C_3 = \dfrac{1}{2} C_4$.

\begin{figure}[h]
	\begin{center}
		\includegraphics[width=8.6cm, height=6.6cm]{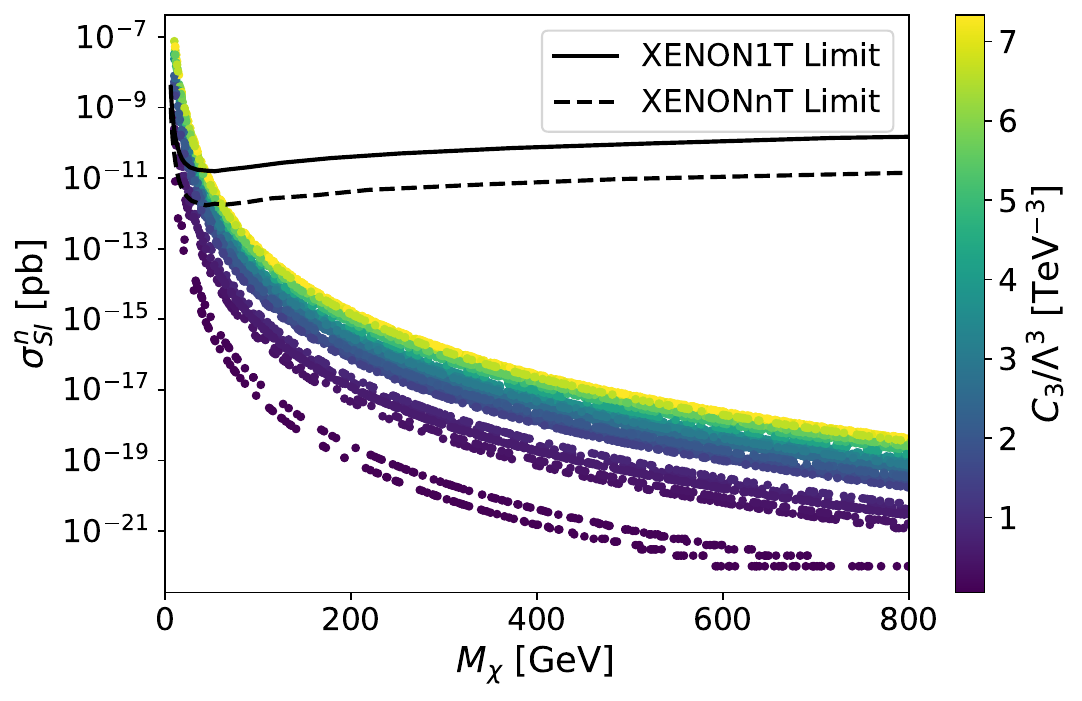}
        \includegraphics[width=8.6cm, height=6.6cm]{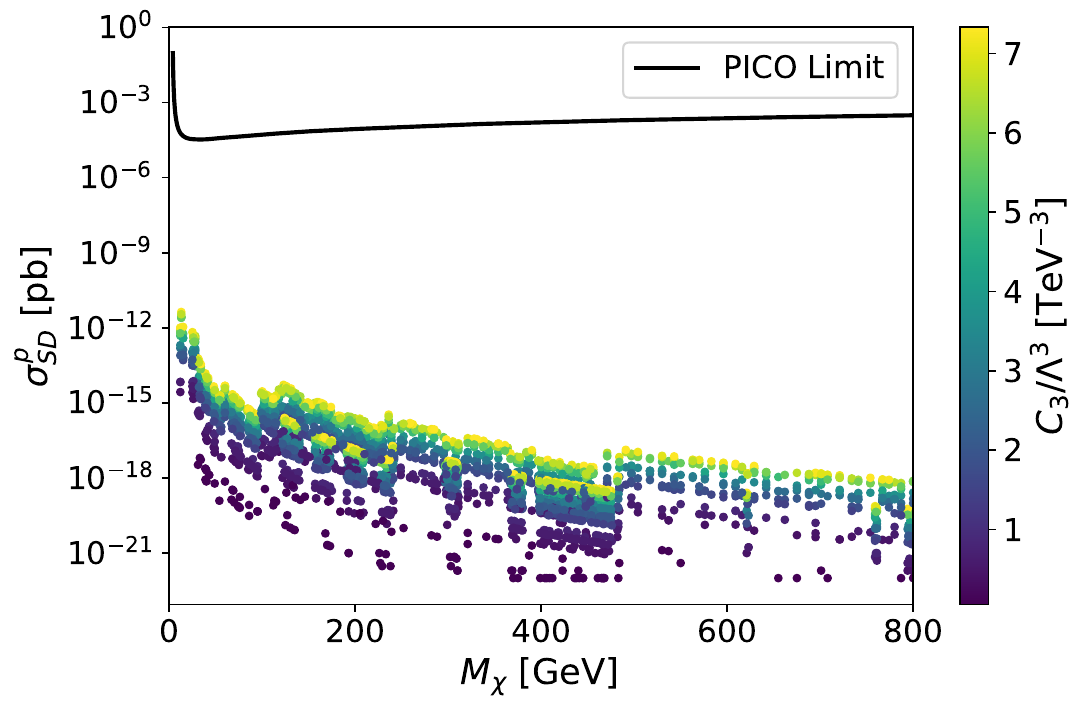}
	 \end{center}
 \caption{Variation of DM-nucleon scattering cross-section with DM mass, $M_{\chi}$, for the combination of $\mathcal{O}_3$ and $\mathcal{O}_4$ in the SI\,(left) and SD\,(right) cases. $\sigma^p_{SD}$ and $\sigma^n_{SI}$ denote DM-proton and DM-neutron scattering cross-sections respectively. The color bar represents the values of $C_3/\Lambda^3$ in TeV$^{-3}$ for both plots and the black solid\,(dashed) line denote the limit\,(projected limit) from the mentioned experiment. In these plots, we consider $C_3 = C_4$.}
 \label{dir_detect_c3c4}
\end{figure}

Let us begin with $C_3 = C_4$. The DM-nucleon SD scattering cross-section arising due to the interactions present in $\mathcal{O}_{3}$ and $\mathcal{O}_{4}$ is very small\,($\sim 10^{-14}$ pb) compared to the present upper limit provided by PICO Collaboration, whereas for the SI case, masses below 45\,(66) GeV are excluded from XENON1T data\,(XENONnT 20\;T-yr projection \cite{xenon3}) for all the values of $C_3/\Lambda^3$\,(or $C_4/\Lambda^3$) considered in our analysis. The SI and SD cross-sections are presented in Fig.~\ref{dir_detect_c3c4} for a range of values of $M_\chi$ and $C_3/\Lambda^3$ \,(or $C_4/\Lambda^3$). The left\,(right) plot\,(colored dots) represents the variation of DM-nucleon SI(SD) scattering cross-section with DM mass. The color gradient shows the values of $C_3/\Lambda^3$ \,(or $C_4/\Lambda^3$) in TeV$^{-3}$. The solid (dashed) line in left panel represents the upper limit\,(projected upper limit) on the corresponding cross-sections from XENON1T\,(XENONnT) at 90\% C.L. In the right panel the solid line denotes the upper limit on DM-nucleon scattering cross-section by PICO experiment at 90\% C.L. Experimental data from LZ Collaboration \cite{lz1, lz2} rules out DM masses below 64 GeV. All the BPs used in our analysis and defined in the next section, satisfy the direct detection constraints. 

\begin{figure}[h]
	\begin{center}
		\includegraphics[width=8.6cm, height=6.6cm]{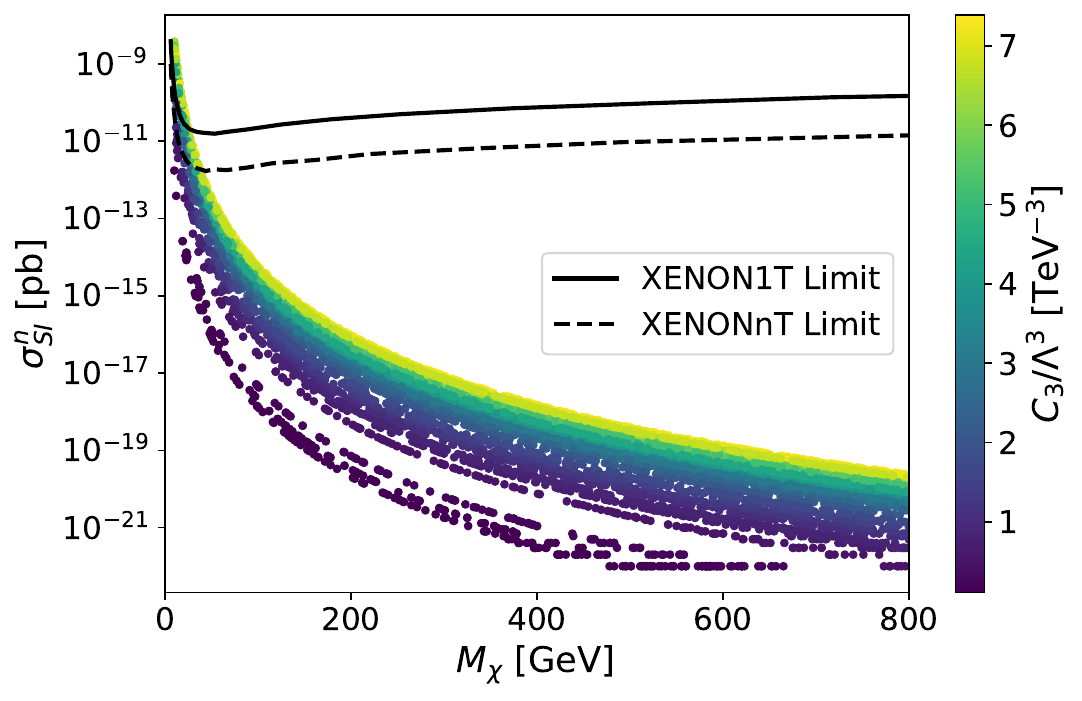}
        \includegraphics[width=8.6cm, height=6.6cm]{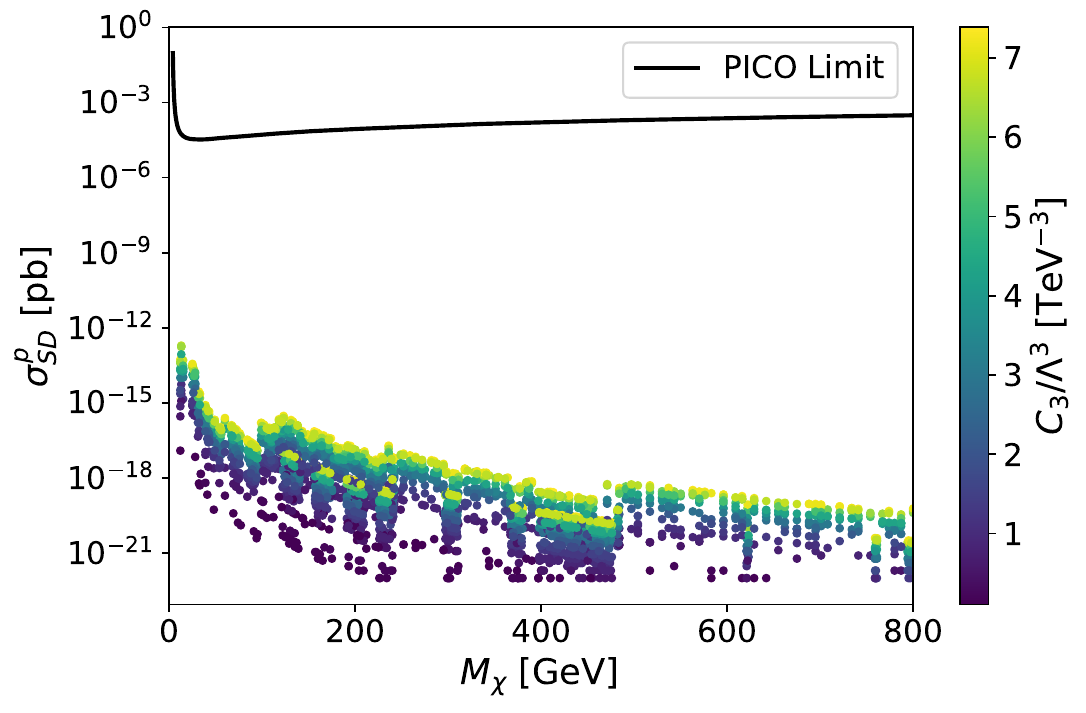}
	 \end{center}
 \caption{Variation of DM-nucleon scattering cross-section with DM mass, $M_{\chi}$, for the combination of $\mathcal{O}_3$ and $\mathcal{O}_4$ in the SI\,(left) and SD\,(right) cases. $\sigma^p_{SD}$ and $\sigma^n_{SI}$ denote DM-proton and DM-neutron scattering cross-sections respectively. The color bar represents the values of $C_3/\Lambda^3$ in TeV$^{-3}$ for both plots and the black solid\,(dashed) line denotes the limit\,(projected limit) from the mentioned experiment. In these plots, we consider $C_3 = \dfrac{1}{2} C_4$.}
 \label{dir_detect_2c3c4}
\end{figure}

 Before discussing the results on relic density, we briefly examine the direct detection cross-sections for $C_3 = \dfrac{1}{2} C_4$. In Fig.~\ref{dir_detect_2c3c4} the left\,(right) plot represents the variation of DM-neutron\,(proton) scattering cross-section as a function of $M_\chi$. In the left plot, the black solid\,(dashed) line denotes 90\% C.L. XENON1T\,(XENONnT projected) limit on DM-neutron scattering cross-section whereas in the right plot, the black solid line describes the PICO limit on DM-proton scattering cross-section. In both cases, the color bar represents the values of $C_3/\Lambda^3$. DM candidates below the mass of 27\,(41) GeV are excluded from XENON1T\,(XENONnT projected) limit.

All the operators in Eqns.~\eqref{efftop1} - \eqref{efftop4} contribute significantly to the relic abundance of the present Universe. In the early Universe, when the temperature was too high\,($T \gg M_{\chi}$), the SM particles were in thermal equilibrium with the DM particles. The rates of creation and annihilation of DM were similar. As the Universe gradually cooled down, at $T < M_{\chi}$, the creation of DM particles from SM particles became kinematically forbidden but DM-DM annihilation to SM particles was still allowed. However, with the expansion of the Universe, $\chi$ became decoupled from the thermal bath. At the time of freeze-out, the rate of annihilation became equal to the Hubble expansion rate and the remnant amount of DM at that time is the relic density of the Universe. One can estimate the relic density solving the Boltzmann equation~\cite{boltzman}. In the following, the relic density measurement is studied considering the interactions described by a combination of $\mathcal{O}_1$ and $\mathcal{O}_2$ and a combination of $\mathcal{O}_3$ and $\mathcal{O}_4$ separately.

\begin{figure}[H]
	\begin{center}
		\includegraphics[width=9cm, height=6.6cm]{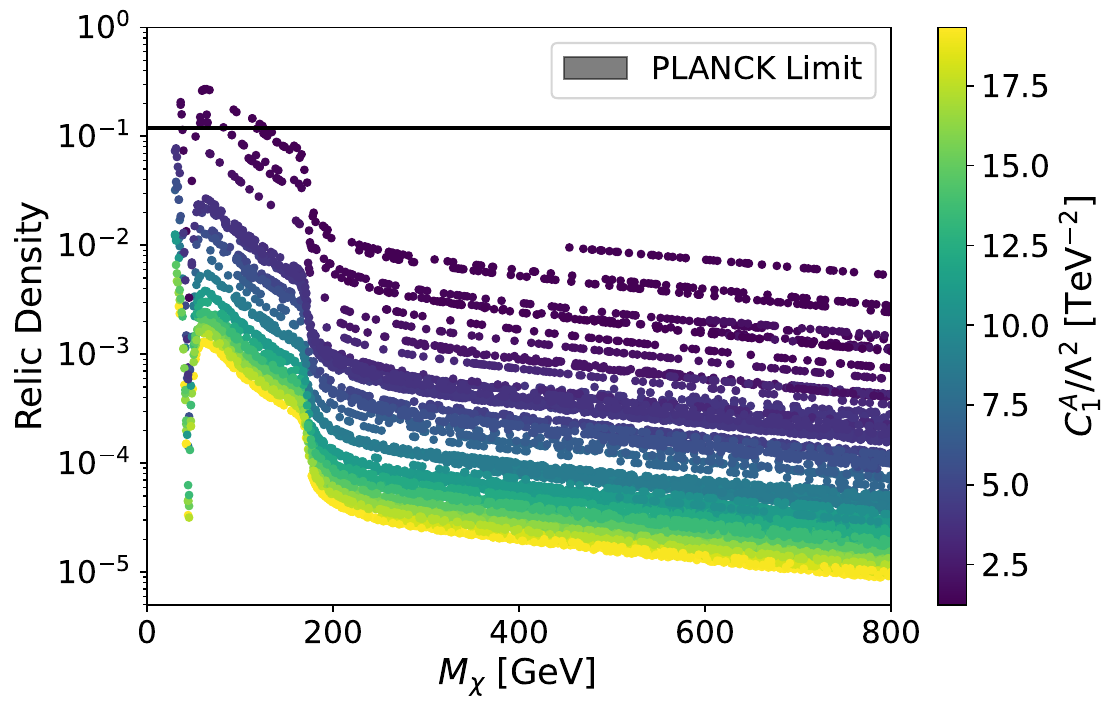}
	\end{center}
 \caption{Variation of relic density with Dark Matter\,(DM) mass, $M_\chi$ for the combination of ${\cal O}_1$ and $\mathcal{O}_2$. The points are allowed by the XENONnT limit on DM-neutron scattering cross-section. The black band shows the PLANCK allowed value of the relic density with $1\sigma$ error.}
 \label{relic_op1}
\end{figure}

Solving the Boltzmann equation one can obtain~\cite{relic}:,
\begin{equation}
    \Omega h^2 \simeq \dfrac{0.1 ~pb}{\langle \sigma v_{rel} \rangle_{eff}}
\end{equation}
where, $h$ is the Hubble constant and $\langle \sigma v_{rel} \rangle_{eff}$ is the thermally averaged effective annihilation cross-section times relative velocity of the DM candidate. This expression clearly shows that a larger annihilation cross-section results in a lower relic density. In this scenario, the DM candidate can annihilate to a pair of SM particles, for example, $ZZ, ~W^+W^-, ~Zh, ~f\bar{f}$ via the effective operators. Fig.~\ref{relic_op1} represents the variation of relic density with $M_\chi$ driven by the interactions, $\mathcal{O}_1$ and $\mathcal{O}_2$ with the color gradient representing the values of the varying Wilson coefficient, $C_1^A$. It is important to note that the values of $|C_2^A|/\Lambda^2$ have been randomly varied in the range $\left [0.0005:0.02 \right]$ TeV$^{-2}$ so that the direct detection limits from XENON are satisfied with the corresponding couplings and masses of the relic particle. Here, the black band denotes the region where the abundance of the relic is within the allowed range of PLANCK, i.e. $\Omega h^2 = 0.12 \pm 0.001$~\cite{Planck}. While considering $\mathcal{O}_1$, a pair of relic particles can dominantly annihilate into a pair of SM fermions or $W^+W^-$ via $Z$ boson as a mediator. Thus, when $M_{\chi}$ approaches $M_Z/2$, the annihilation cross-section is very large resulting a very small value of relic density. This explains the dip near $M_{\chi} \simeq M_Z/2$ in the plot in Fig.~\ref{relic_op1}. The sudden drop in the abundance of the relics around $M_{\chi} = m_t$ could be explained by the increase of $\langle \sigma v_{rel} \rangle$ due to the opening of the $\chi \bar \chi \rightarrow t \bar t$ channel. In general, for larger values of the Wilson coefficients, the cross section of annihilation increases and the density of the relics decreases in turn. This explains the color gradient of the plots. The BPs of our analysis also satisfy the PLANCK limit on the current relic abundance of the Universe.

 The operators ${\cal O}_3  {~\rm and} ~{\cal O}_4$ also contribute to the relic density. In Fig.~\ref{relic_op34},
variation of relic density with $\chi$, has been presented for a range of values of $C_3$. Two choices of $C_4$ have been assumed, namely $C_3 = C_4$\,(a) and $C_3 = \dfrac{1}{2}C_4$\,(b, c). Qualitatively, these two choices will not reveal any new feature of the inherent dynamics of the operators. However, for a fixed value of $C_3$, $\langle \sigma v_{rel} \rangle$ is slightly higher for the choice $C_3 = \dfrac{1}{2} C_4$, which in turn results in a lower value of the relic density. This is clearly evident when we compare Fig.~\ref{relic_op34}(a) with Fig.~\ref{relic_op34}(b) and Fig.~\ref{relic_op34}(c).

\begin{figure}[H]
	\centering
		\subfloat[]{\includegraphics[width=8.6cm, height=6.6cm]{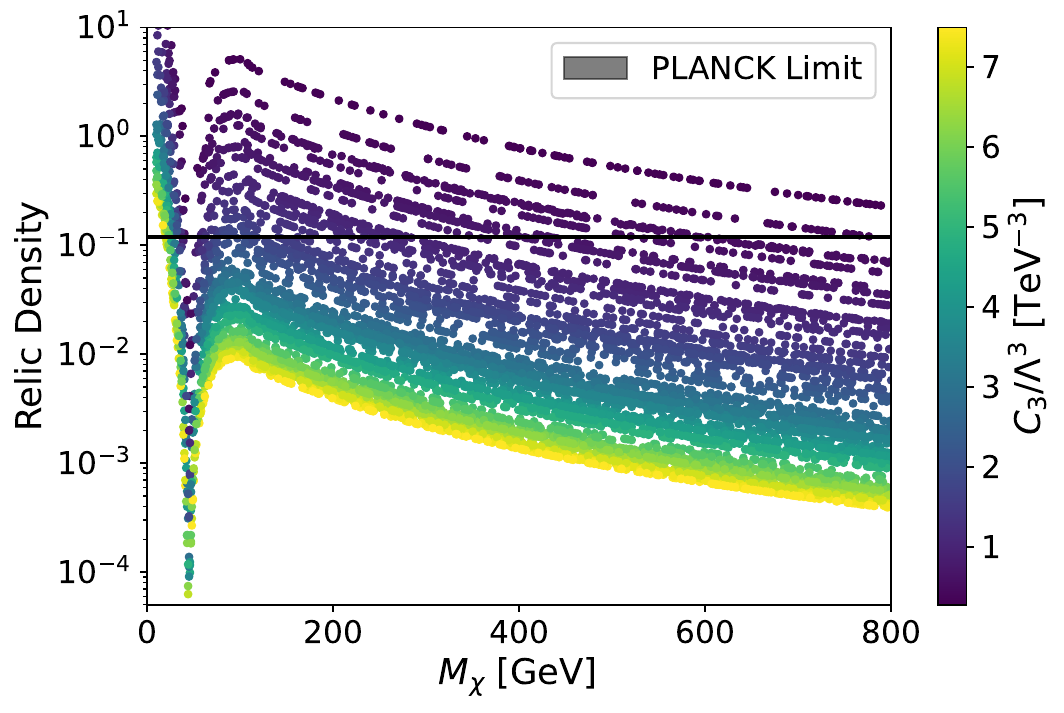}}
        \phantomcaption
\end{figure}   
\begin{figure}[H]
\ContinuedFloat
\centering    
        \subfloat[]{\includegraphics[width=8.6cm, height=6.6cm]{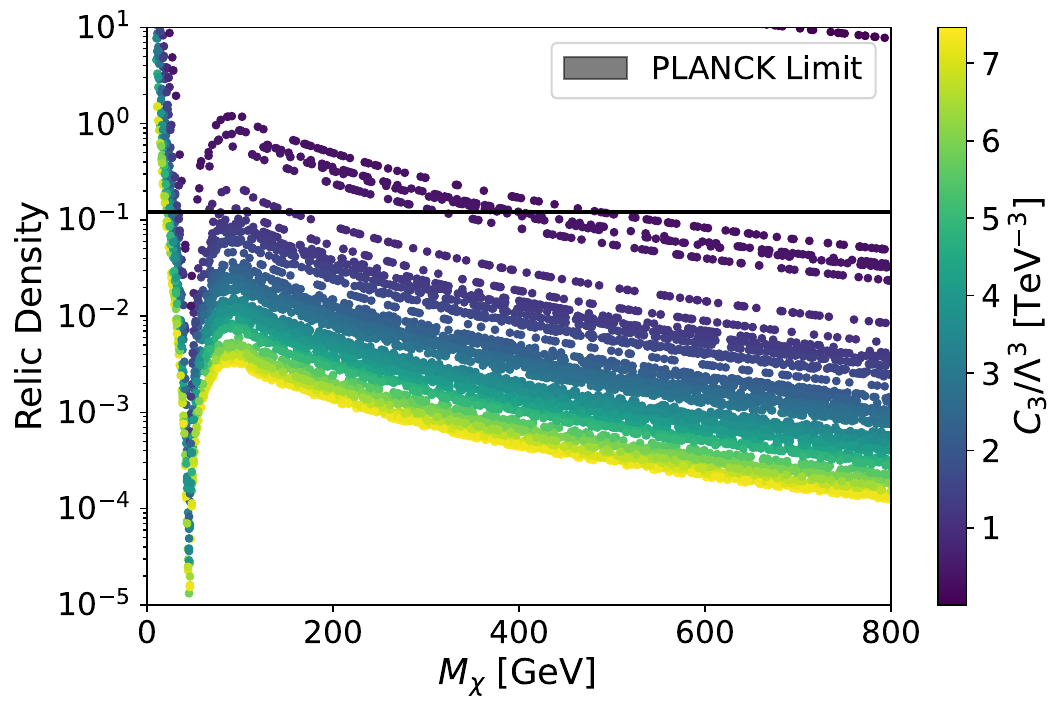}}
        \subfloat[]{\includegraphics[width=8.6cm, height=6.6cm]{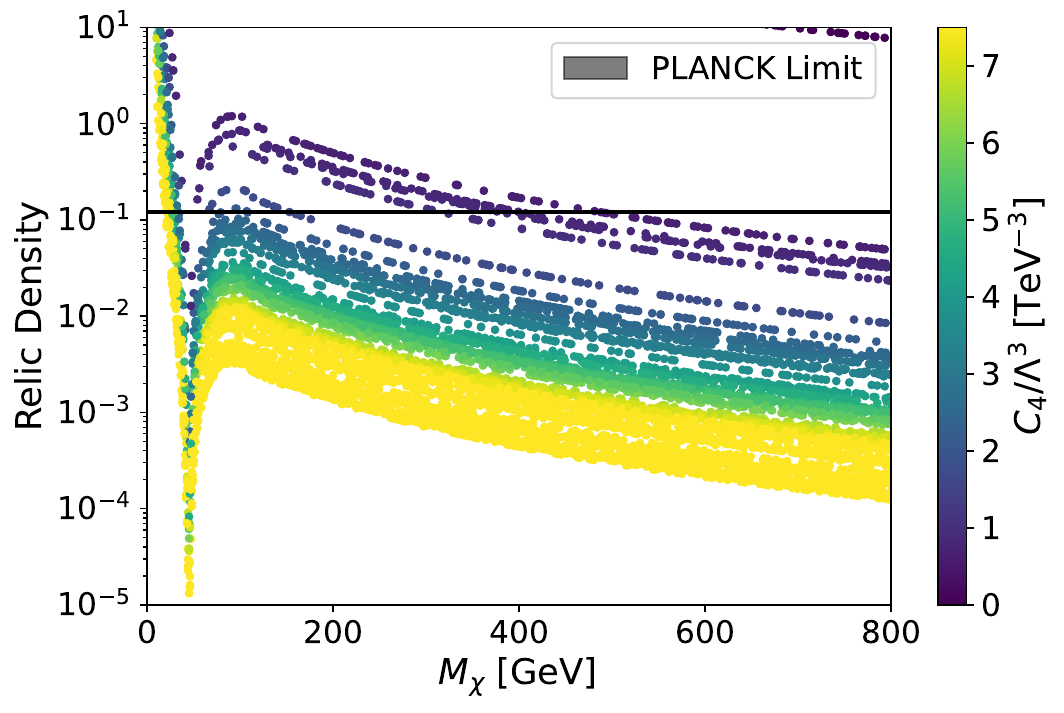}}
 \caption{Variation of the relic density with DM mass, $M_{\chi}$, for the combination of ${\cal O}_3$ and ${\cal O}_4$. For plot\,(a), $C_3=C_4$, while for plot (b) and (c), $C_3=\dfrac{1}{2} C_4$. The color bar in (a) and (b) represents the values of $C_3/\Lambda^3$ in TeV$^{-3}$, whereas in (c) the color bar denotes the values of $C_4/\Lambda^3$ in TeV$^{-3}$. The black band shows the PLANCK allowed value of the relic density with $1\sigma$ error.}
 \label{relic_op34}
\end{figure}

Before concluding this section, we briefly discuss the choice of parameters, \( M_\chi \) and the Wilson coefficients, used in this study. The annihilation cross-section of the relic particle, which is used to estimate the relic density, and the cross-section for pair production of \( \chi \) with a Higgs boson, both involve scattering processes driven solely by the effective interactions discussed earlier. Consequently, these cross-sections scale with either \( \frac{C_1^2}{\Lambda^4} \) or a specific combination of \( \frac{C_3^2}{\Lambda^6} \) and \( \frac{C_4^2}{\Lambda^6} \) for a fixed \( M_\chi \). The interaction rate decreases monotonically with increasing relic mass in the final state, as there are no other particles in the spectrum whose decay or threshold production could alter the monotonic behavior of the cross-sections. As a result, our choice of BPs sufficiently explores the entire parameter space. From Figs.~\ref{dir_detect_o1}, \ref{dir_detect_c3c4}, \ref{dir_detect_2c3c4}, \ref{relic_op1}, and \ref{relic_op34}, one can identify parameter values that satisfy both relic density and direct detection constraints.

\section{Collider Search for Dark Matter} \label{sec3}

In general, production of the relic particle, $\chi$, at the LHC will result in missing transverse energy\,(momentum). For a conclusive signature of $\chi$ production in an event, one must tag some objects which could be easily identified in the detector. In view of the effective interactions\,(Eqns.~\ref{efftop1},~\ref{efftop3} and \ref{efftop4}), $\chi$ can be pair produced along with the SM Higgs boson in quark-antiquark annihilation, as shown in the Feynman diagrams in Fig.~\ref{feynman}. 
The signal events consist of two b-jets coming from the Higgs boson decay along with a large missing transverse energy\,(MET). 
Two signal configurations with similar final states are studied: one primarily described by the operator $\mathcal{O}_1$ and the other as a combination of operators $\mathcal{O}_3$ and $\mathcal{O}_4$. It is worth mentioning that $\mathcal{O}_2$ does not play any role in the collider signature of mono-Higgs search for $\chi$. 

SM processes having similar final states will act as the background. Hadronic, leptonic and semileptonic decays of $t\bar{t}$, ZZ and Zh processes are the major sources of background for this analysis.

\begin{figure}[H]
	\begin{center}
\begin{tikzpicture}
	\begin{feynman}
		\vertex(a) at (-2,-2) {\(q\)};
		\vertex(b) at (0,0);
		\vertex(c) at (-2,2) {\(\bar{q}\)};
		\vertex[blob] (d) at (2,0) {\contour{white}{}};
		\vertex(h1) at (3,1);
		\vertex(f3) at (3.7,2) {\(b\)};
		\vertex(f4) at (4,0.8) {\(b\)};
		\vertex(f1) at (4,0) {\(\chi\)};
		\vertex(f2) at (4,-2) {\(\chi\)};
		\diagram* {(a) -- [fermion] (b), (b) -- [fermion] (c),(b) -- [boson,edge label'=\(Z\)] (d),(d) -- [scalar, edge label'=\(h\)] (h1),(d) -- [anti fermion] (f2),(d) -- [fermion] (f1),(h1) -- [fermion] (f3),(h1) -- [anti fermion] (f4)};
	\end{feynman}
\end{tikzpicture}
\begin{tikzpicture}
	\begin{feynman}
		\vertex(a) at (-2,-2) {\(q\)};
		\vertex(b) at (0,0);
		\vertex(c) at (-2,2) {\(\bar{q}\)};
		\vertex(d) at (2,0);
		\vertex(h1) at (3,1);
		\vertex[blob] (b1) at (3,-1) {\contour{white}{}};
		\vertex(f1) at (4,-0.2) {\(\chi\)};
		\vertex(f2) at (4,-2) {\(\chi\)};
		\vertex(f3) at (4,0.2) {\(b\)};
		\vertex(f4) at (4,2) {\(b\)};
		\diagram* {(a) -- [fermion] (b), (b) -- [fermion] (c),(b) -- [boson,edge label'=\(Z\)] (d),(d) -- [scalar, edge label'=\(h\)] (h1),(d) -- [boson, edge label'=\(Z\)] (b1),(b1) -- [fermion] (f1),(b1) -- [anti fermion] (f2),(h1) -- [fermion] (f3),(h1) -- [anti fermion] (f4)};
	\end{feynman}
\end{tikzpicture}
\end{center}
\caption{Feynman diagrams describing the effective interactions mentioned in Eqns.~\ref{efftop1}, \ref{efftop3} and \ref{efftop4}.}
	\label{feynman}
\end{figure}

 The signal events have been generated by interfacing \texttt{FeynRules} output with the event generator, $\texttt{MadGraph5\_aMC@NLO}$~\cite{Alwall_2014}, requiring a minimum transverse momentum\,($p_{T}$) of $5~GeV$ for the b quarks. The SM backgrounds are also simulated using $\texttt{MadGraph5\_aMC@NLO}$ having the same cut on $p_{T}$ of b quark. For both signal and background samples, \texttt{Pythia8}~\cite{bierlich2022comprehensiveguidephysicsusage} is used for parton showering, hadronisation and decay, followed by detector simulation in \texttt{Delphes3.5.0}~\cite{de_Favereau_2014}. The input card with CMS detector setup for HL-LHC has been used. A sample of $10^5$ minimum bias events are used for PU mixing with an average PU of 200.  
  The BPs for both the signal configurations are set by varying \(M_{\chi}\) and the Wilson coefficients as shown in Table~\ref{tab:relic_density_xsec_C1C2} and \ref{tab:relic_density_xsec_C3_C4}. For the signal corresponding to operator \(\mathcal{O}_1\), six BPs are considered for each of the following two sets of Wilson coefficient values:  
\begin{itemize}
    \item \textbf{Set A:} \(C_1^A/\Lambda^2 = 8.75\) TeV\(^{-2}\), \(|C_2^A|/\Lambda^2 = 8.6 \times 10^{-3}\) TeV\(^{-2}\)  
    \item \textbf{Set B:} \(C_1^A/\Lambda^2 = 15\) TeV\(^{-2}\), \(|C_2^A|/\Lambda^2 = 1.474 \times 10^{-2}\) TeV\(^{-2}\)  
\end{itemize}  

Similarly, for the signal arising from the combination of operators \(\mathcal{O}_3\) and \(\mathcal{O}_4\), six BPs are chosen for each of the following two sets of Wilson coefficient values:  
\begin{itemize}
    \item \textbf{Set C:} \(C_3/\Lambda^3 = C_4/\Lambda^3 = 5\) TeV\(^{-3}\) 
    \item \textbf{Set D:} \(C_3/\Lambda^3 = \frac{1}{2} C_4/\Lambda^3 = 2\) TeV\(^{-3}\) 
\end{itemize}
The Wilson coefficients\,($C/\Lambda^{2(3)}$) are consistent with the limits on relic density and direct detection cross-section of relic-nucleon scattering. At this point, it is also worth mentioning that these BPs also satisfy the upper limit on thermally averaged annihilation cross-section times relative velocity of the annihilating relic particles ($v_{rel}$) provided by FermiLAT in $b\bar{b}$ channel\,(from 4 years and 6 years of data taking). The relic density and cross section times branching fraction for $h\rightarrow b\bar{b}$ for all the different BPs are listed in Tables\,\ref{tab:relic_density_xsec_C1C2} and \ref{tab:relic_density_xsec_C3_C4}. It should also be mentioned that both $\mathcal{O}_1$ and combination of $\mathcal{O}_3$ and $\mathcal{O}_4$ can induce Higgs\,(Z) invisible decay through $h Z \chi \chi$ interaction for relic masses less than $M_h/2\,(M_Z/2)$. In such a case, the limits of Higgs\,($Z$) invisible decay rate would be applicable on the Wilson coefficients. However, for the BPs used in our analysis such a decay of the Higgs\,($Z$) boson is kinematically forbidden.

\begin{table}[H]
    \centering
    \renewcommand{\arraystretch}{1.5}
    \begin{tabular}{|c|c|c|c|}
    \hline 
     & $M_{\chi} \,(\text{GeV})$ &  Relic Density & $\sigma_{prod} \times BR \,(\text{fb})$ \\ [2.0ex]
    \hline \hline
    BP1A & $90$ &  $3.76 \times 10^{-3}$ & 0.521 \\ 
    \hline
    BP1B & $90$ &  $1.37 \times 10^{-3}$ & 1.531 \\ 
    \hline \hline
    BP2A & $100$ &  $3.08 \times 10^{-3}$ & 0.456\\ 
    \hline
    BP2B & $100$ &  $1.11 \times 10^{-3}$ & 1.336 \\ 
    \hline \hline 
    BP3A & $150$ &  $1.47 \times 10^{-3}$ & 0.254 \\ 
    \hline
    BP3B & $150$ &  $5.32 \times 10^{-4}$ & 0.748 \\ 
    \hline \hline 
    BP4A & $200$ & $4.59 \times 10^{-4}$ & 0.154 \\ 
    \hline
    BP4B & $200$ &  $1.64 \times 10^{-4}$ & 0.453 \\ 
    \hline \hline 
    BP5A & $250$ &  $3.1 \times 10^{-4}$ & 0.098 \\ 
    \hline
    BP5B & $250$ &  $1.11 \times 10^{-4}$ & 0.287 \\ 
    \hline \hline 
    BP6A & $300$ &  $2.4 \times 10^{-4}$ & 0.064 \\ 
    \hline
    BP6B & $300$ &  $8.60 \times 10^{-5}$ & 0.189 \\ 
    \hline \hline       
    \end{tabular}
    \caption{Relic density and cross section times branching fraction for different $M_{\chi}$ and two values of Wilson coefficients 
    $C_1$ and $C_2$. For the Benchmark Points\,(BPs) marked with \textbf{A}, $C_1^A/\Lambda^2 = 8.75 ~\rm TeV^{-2}$ and $|C_2^A|/\Lambda^2 = 8.6 \times 10^{-3} ~\rm TeV^{-2}$, while $C_1^A/\Lambda^2 = 15 ~\rm TeV^{-2}$ and $|C_2^A|/\Lambda^2 = 1.474 \times 10^{-2} ~\rm TeV^{-2}$ for the BPs marked with \textbf{B}.}
    \label{tab:relic_density_xsec_C1C2}
\end{table}

\begin{table}[H]
    \centering
    \renewcommand{\arraystretch}{1.5}
    \begin{tabular}{|c|c|c|c|}
    \hline 
     & $M_{\chi} \,(\text{GeV})$ &  Relic Density & $\sigma_{prod} \times BR \,(\text{fb})$ \\ [2.0ex]
    \hline \hline
    BP1C & $90$ & $1.97 \times 10^{-2}$ & 1.862 \\ 
    \hline
    BP1D & $90$ & $4.04 \times 10^{-2}$ & 0.934 \\ 
    \hline\hline  
    BP2C & $100$ & $2.00 \times 10^{-2}$ & 1.822 \\ 
    \hline
    BP2D & $100$ & $4.02 \times 10^{-2}$ & 0.899 \\ 
    \hline\hline  
    BP3C & $150$ & $1.21 \times 10^{-2}$ & 1.615 \\ 
    \hline
    BP3D & $150$ & $2.39 \times 10^{-2}$ & 0.809 \\ 
    \hline \hline 
    BP4C & $200$ & $8.56 \times 10^{-3}$ & 1.414 \\ 
    \hline
    BP4D & $200$ & $1.66 \times 10^{-2}$ & 0.730 \\
    \hline\hline  
    BP5C & $250$ & $6.27 \times 10^{-3}$ & 1.226 \\
    \hline
    BP5D & $250$ & $1.20 \times 10^{-2}$ & 0.628 \\ 
    \hline\hline  
    BP6C & $300$ & $4.75 \times 10^{-3}$ & 1.059 \\ 
    \hline
    BP6D & $300$ & $9.09 \times 10^{-3}$ & 0.546 \\
   \hline \hline
    \end{tabular}
    \caption{Relic density and cross section times branching fraction for different $M_{\chi}$ and two values of Wilson coefficients 
    $C_3$ and $C_4$. For the Benchmark Points\,(BPs) marked with \textbf{C}, $C_3/\Lambda^3 = C_4/\Lambda^3 = 5 ~\rm TeV^{-3}$, while $C_3/\Lambda^3 = 2 ~\rm TeV^{-3}$ and $C_4/\Lambda^3 = 4 ~\rm TeV^{-3}$ for the BPs marked with \textbf{D}.}
    \label{tab:relic_density_xsec_C3_C4}
\end{table}
\pagebreak
The production of a $\chi$ pair in association with a $Z$-boson is also possible and can be driven by the same set of operators. The $\chi \bar \chi Z$ production mediated by a Higgs boson from a gluon-gluon initial state results in a $b \bar b$ + MET final state when the $Z$ boson decays to a pair of $b$ quarks. Despite the high gluon luminosity available at the LHC, the cross section for $\chi \bar \chi Z$ production is small compared to $\sigma\,(\chi \bar  \chi h)$ due to the smallness of the effective $hgg$ coupling. For example, for $M_{\chi} = 100\,(200) ~\rm GeV$, the production rate for $\chi \bar \chi Z\,(b \bar b)$ is $0.23\,(0.035) ~\rm fb$ for ${\cal O}_1$ with $C_1^A/\Lambda^2=15$ TeV$^{-2}$ and $0.07\,(0.036) ~\rm fb$ for the combination of ${\cal O}_3$ and ${\cal O}_4$\,(with $C_3=C_4$). We have not included the contribution from $\chi \bar \chi Z$ production in our analysis.

 Hadronic, leptonic and semileptonic decays of the $t\bar{t}$ process are the dominant sources of background for our proposed signal, each producing a $b\bar{b}$ pair in the final state with additional jets, leptons and/or MET. Hadronic decays of $t\bar{t}$ with the highest branching have MET signature due to jet energy mis-measurements. The semileptonic and leptonic decays have MET coming from the undetected neutrino\,(s). A lepton veto at the event level helps in reducing the semileptonic and leptonic background. The other major backgrounds in the analysis are $ZZ$ and $Zh$ processes where one $Z$ decays into a pair of neutrinos, while the other leg\,($Z/h$) decays to $b\bar{b}$. The cross-section times branching ratio of the background processes are listed in Table~\ref{background}.
\begin{table}[H]
    \centering
    \renewcommand{\arraystretch}{1.5}
    \begin{tabular}{|c|c|}
	\hline  
        $SM Process$  &  $\sigma_{prod} \times BR \,(\text{pb})$ \\ 
	\hline \hline
	$t \bar t$ Hadronic\,(ttH) &  $131.381$  \\
	\hline
 	$t \bar t$ Leptonic\,(ttL) & $20.453$  \\     
        \hline
        $t \bar t$ SemiLeptonic\,(ttS) & $52.423$  \\
        \hline  
 	$ZZ$ & $0.597$ \\
        \hline
 	$Zh$ & $0.104$  \\
        \hline       
    \end{tabular}
    \caption{Cross-sections times branching ratio of the SM background processes.}
    \label{background}
\end{table}

\subsection{Event Selection and Analysis Strategy} \label{sec3.1}
 All the physics objects used in this analysis are reconstructed from the simulated detector response in \texttt{Delphes}. The jets are reconstructed using the anti-$k_{t}$ jet clustering algorithm~\cite{Cacciari_2008} from the \texttt{FastJet} package~\cite{Cacciari_2012}, with a jet radius parameter of 0.4. For b-tagging, the medium working point provided by \texttt{Delphes} has been used in this analysis. A veto is applied to remove events having isolated leptons with $p_{T} > 10 ~GeV$. Jets passing the Pileup Per Particle Identification\,(PUPPI) algorithm~\cite {kreis2018particle} in \texttt{Delphes} with $p_{T} > 35~GeV$ are selected.
 
\begin{figure}[!htp]
    \begin{minipage}[c]{0.5\linewidth}%
        \vspace{0pt}%
        \centering%
        \subfloat[$p_{T}$ of leading b-jet]{%
        \includegraphics[width=\textwidth]{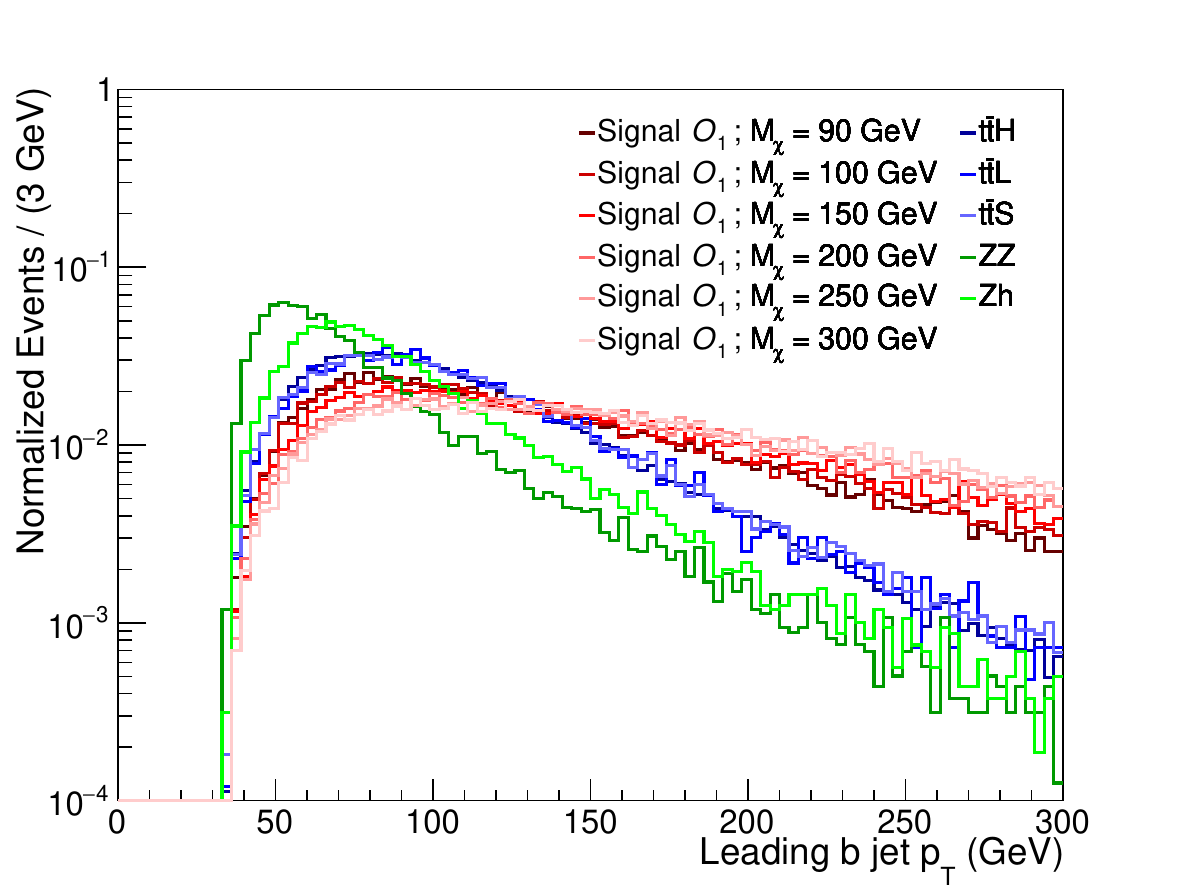}  
        }%
    \end{minipage}
    \begin{minipage}[c]{0.5\linewidth}%
        \vspace{0pt}%
        \centering%
        \subfloat[$p_{T}$ of sub-leading b-jet]{%
         \includegraphics[width=\textwidth]{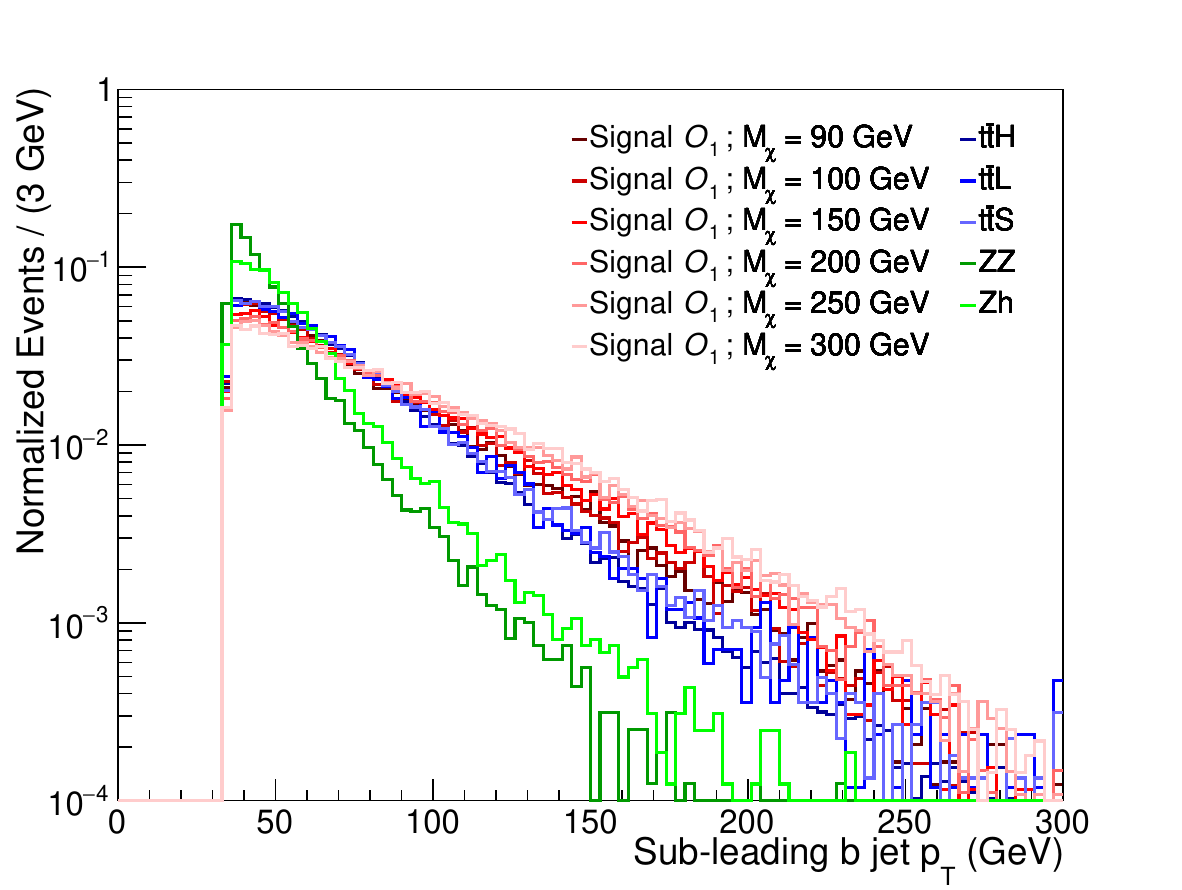} 
        }%
    \end{minipage}
    \begin{minipage}[c]{0.5\linewidth}%
        \vspace{0pt}%
        \centering%
        \subfloat[$\Delta{R}$ between the leading and sub-leading b-jets]{%
         \includegraphics[width=\textwidth]{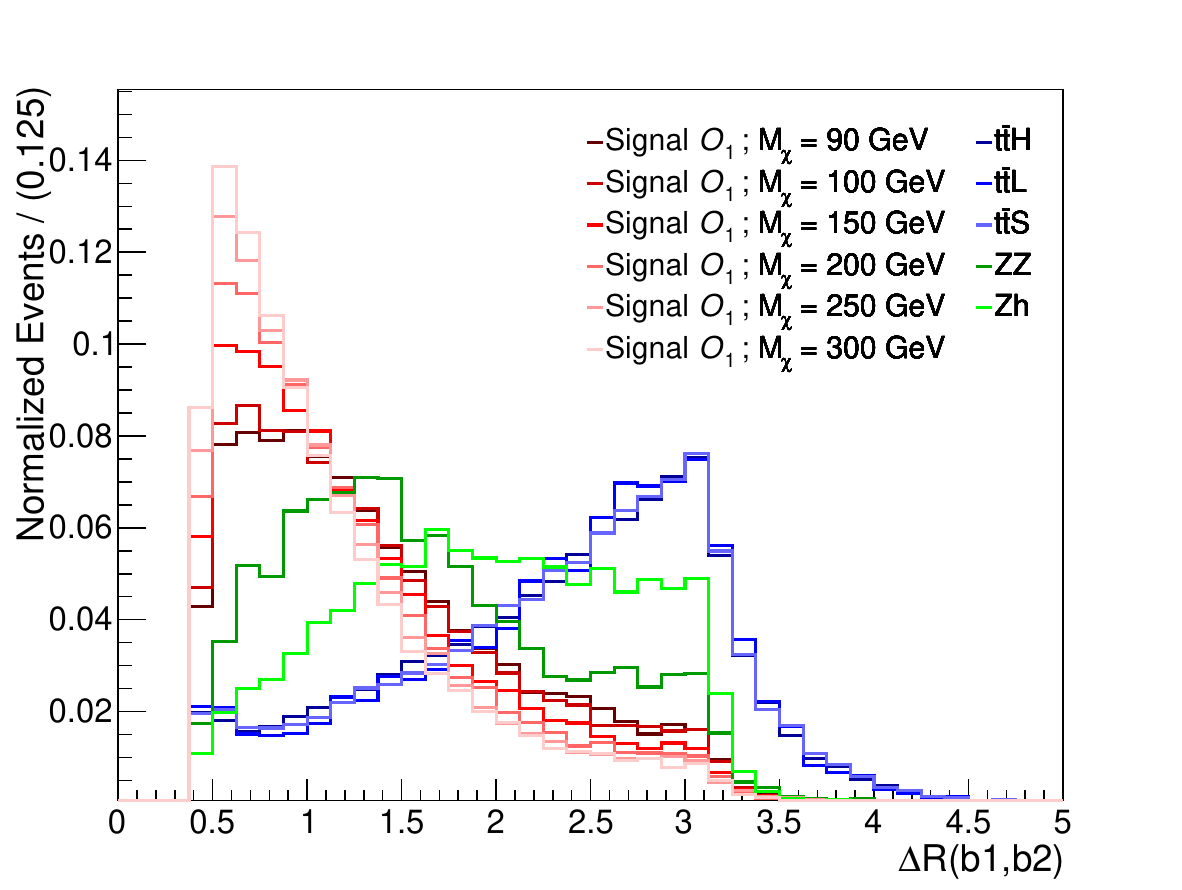} 
        }%
    \end{minipage}
        \begin{minipage}[c]{0.5\linewidth}%
        \vspace{0pt}%
        \centering%
        \subfloat[Invariant mass of the two highest $p_{T}$ b-jets]{%
         \includegraphics[width=\textwidth]{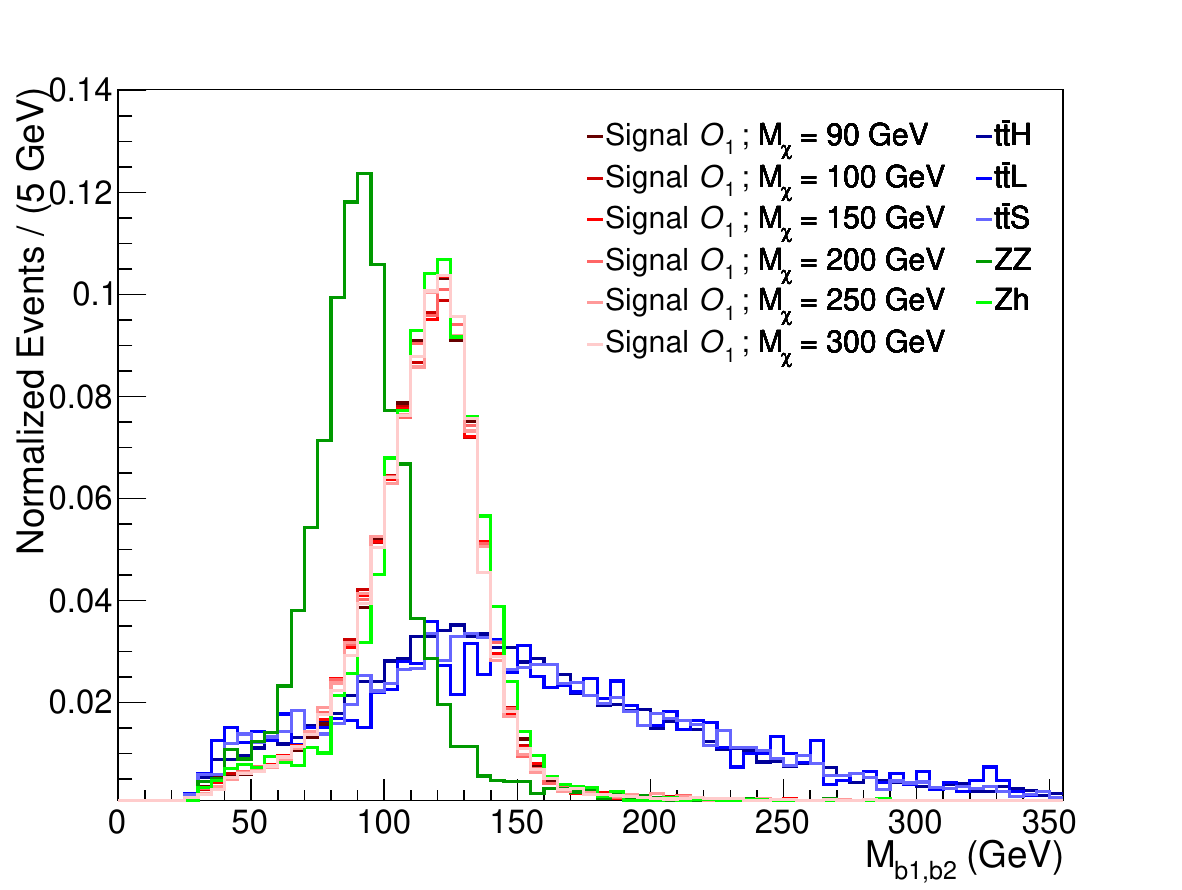}  
        }%
    \end{minipage}
    \begin{minipage}[c]{0.5\linewidth}%
        \vspace{0pt}%
        \centering%
        \subfloat[$\not\!\! E_{T}$]{%
         \includegraphics[width=\textwidth]{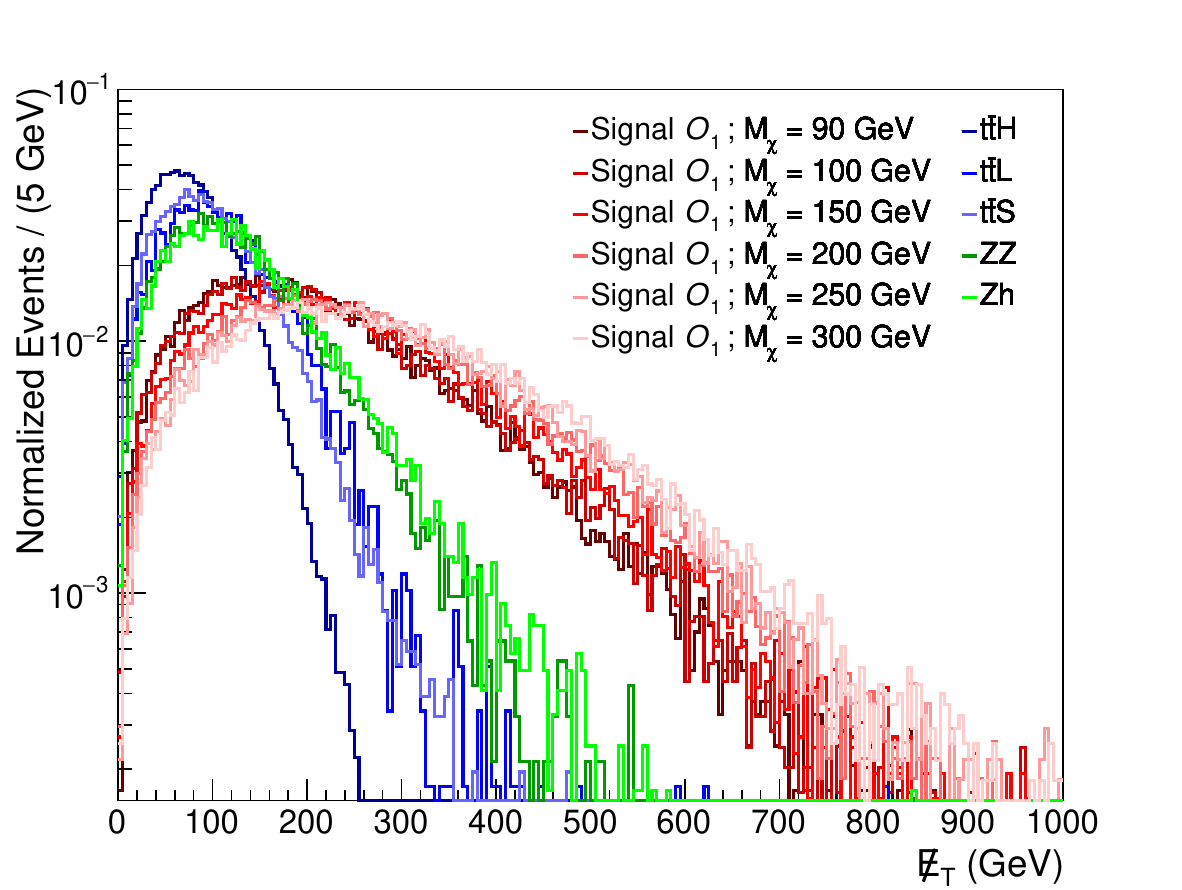}  
        }%
    \end{minipage}
    \begin{minipage}[c]{0.5\linewidth}%
        \vspace{0pt}%
        \centering%
        \subfloat[$\Delta{\phi}$ between the leading b-jet and $\not\!\! E_{T}$]{%
         \includegraphics[width=\textwidth]{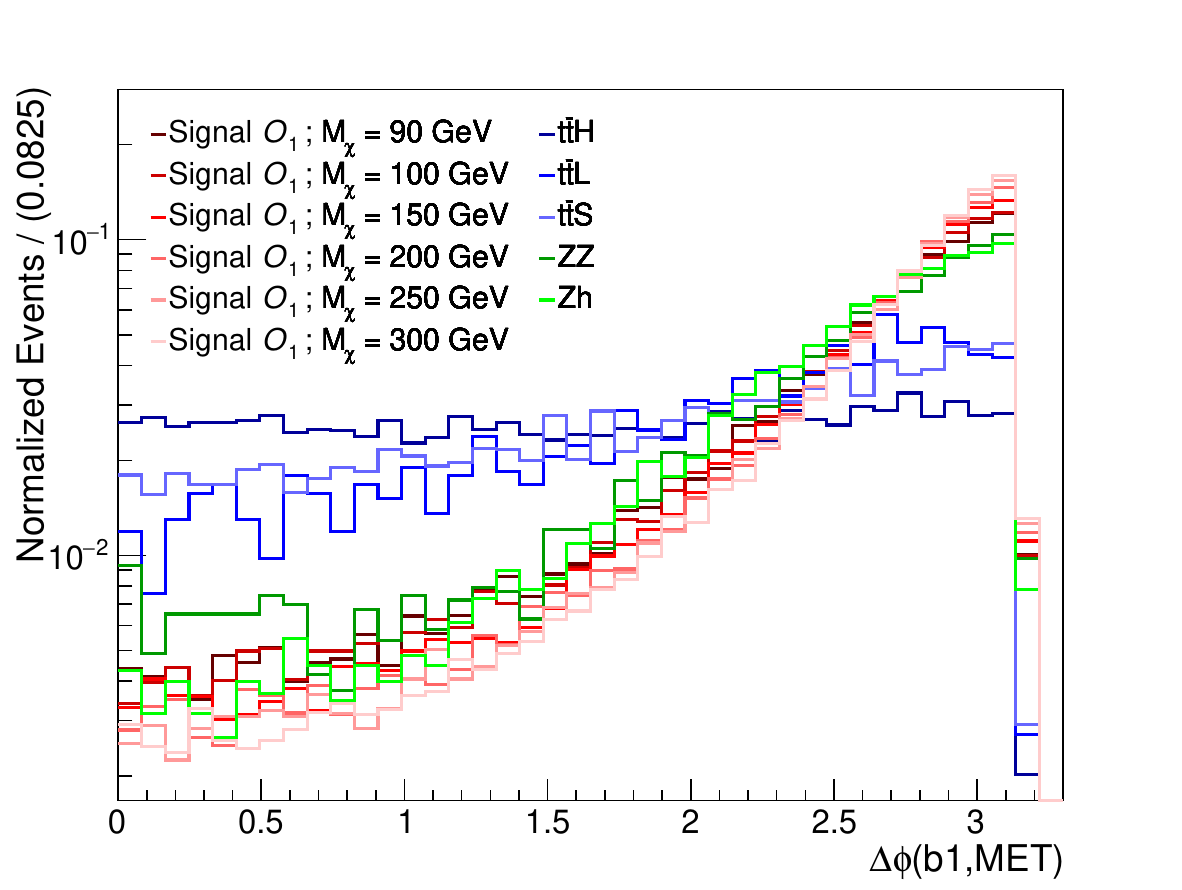} 
        }%
    \end{minipage}    
    \caption{Distributions of key kinematic variables for signals corresponding to $\mathcal{O}_1$ with $C_1^A/\Lambda^2=15$ TeV$^{-2}$ and background processes, all normalized to unity.}
    \label{fig:Kinematics1}
\end{figure}

 We first compare kinematic variables which provide effective separation between signal and background. Distributions of the key kinematic variables for signal\,($\mathcal{O}_1$) and background processes, all normalized to unity, are shown in Figs.~\ref{fig:Kinematics1}\,(a)-(f). The $p_{T}$ distribution of the leading and sub-leading b-tagged jets are harder for signal as shown in Figs.~\ref{fig:Kinematics1}\,(a) and~\ref{fig:Kinematics1}\,(b), respectively.  Fig.~\ref{fig:Kinematics1}\,(c) shows the angular separation, $\Delta{R}$, between the leading and sub-leading b-jets. The Higgs boson recoils against the DM candidates leading to boosted b-jets in the final state. The b-jets from signal should have an invariant mass peaking at the mass of the Higgs boson which is visible in Fig.~\ref{fig:Kinematics1}\,(d). The b-jets in $Zh$ process, also coming from the Higgs decay has an invariant mass distribution similar to signal. Fig.~\ref{fig:Kinematics1}\,(e) shows the distribution of MET\,($\not\!\! E_{T}$). It can be seen that the signal has a large $\not\!\! E_{T}$ in comparison to the background processes. Fig.~\ref{fig:Kinematics1}\,(f) shows the $\Delta\phi$ between the leading b-jet and $\not\!\! E_{T}$.

\begin{figure}[htp]
    \begin{minipage}[c]{0.5\linewidth}%
        \vspace{0pt}%
        \centering%
        \subfloat[$p_{T}$ of leading b-jet]{%
        \includegraphics[width=\textwidth]{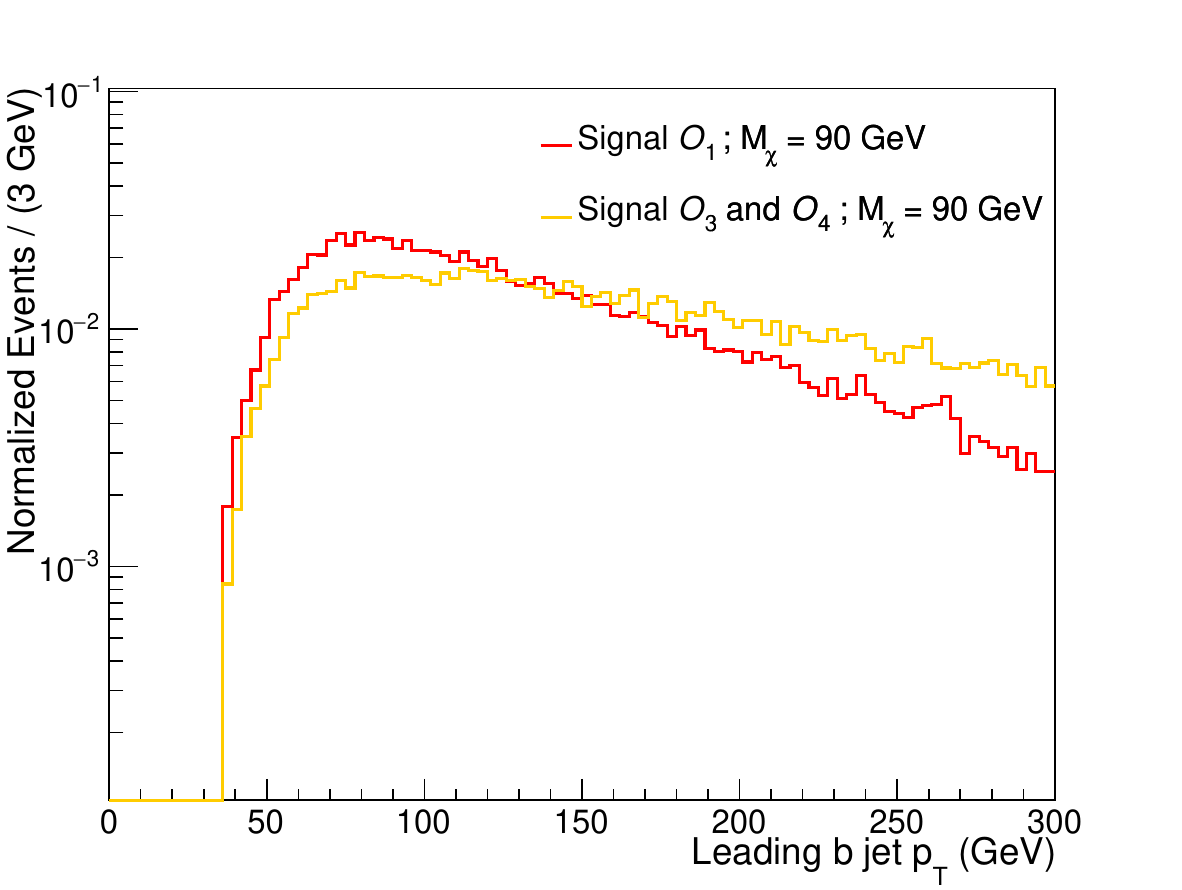}  
        }%
    \end{minipage}
    \begin{minipage}[c]{0.5\linewidth}%
        \vspace{0pt}%
        \centering%
        \subfloat[$p_{T}$ of sub-leading b-jet]{%
         \includegraphics[width=\textwidth]{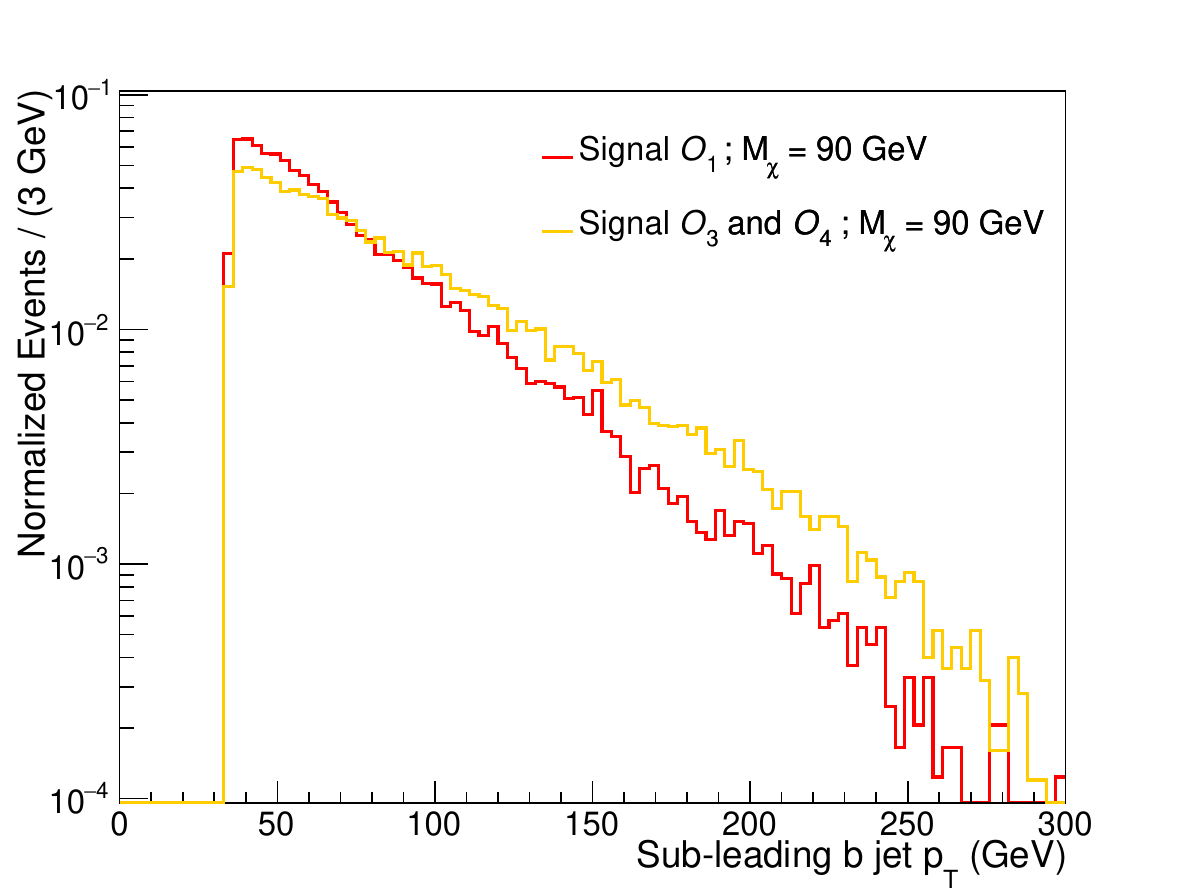}  
        }%
    \end{minipage}
    \begin{minipage}[c]{0.5\linewidth}%
        \vspace{0pt}%
        \centering%
        \subfloat[$\not\!\! E_{T}$]{%
         \includegraphics[width=\textwidth]{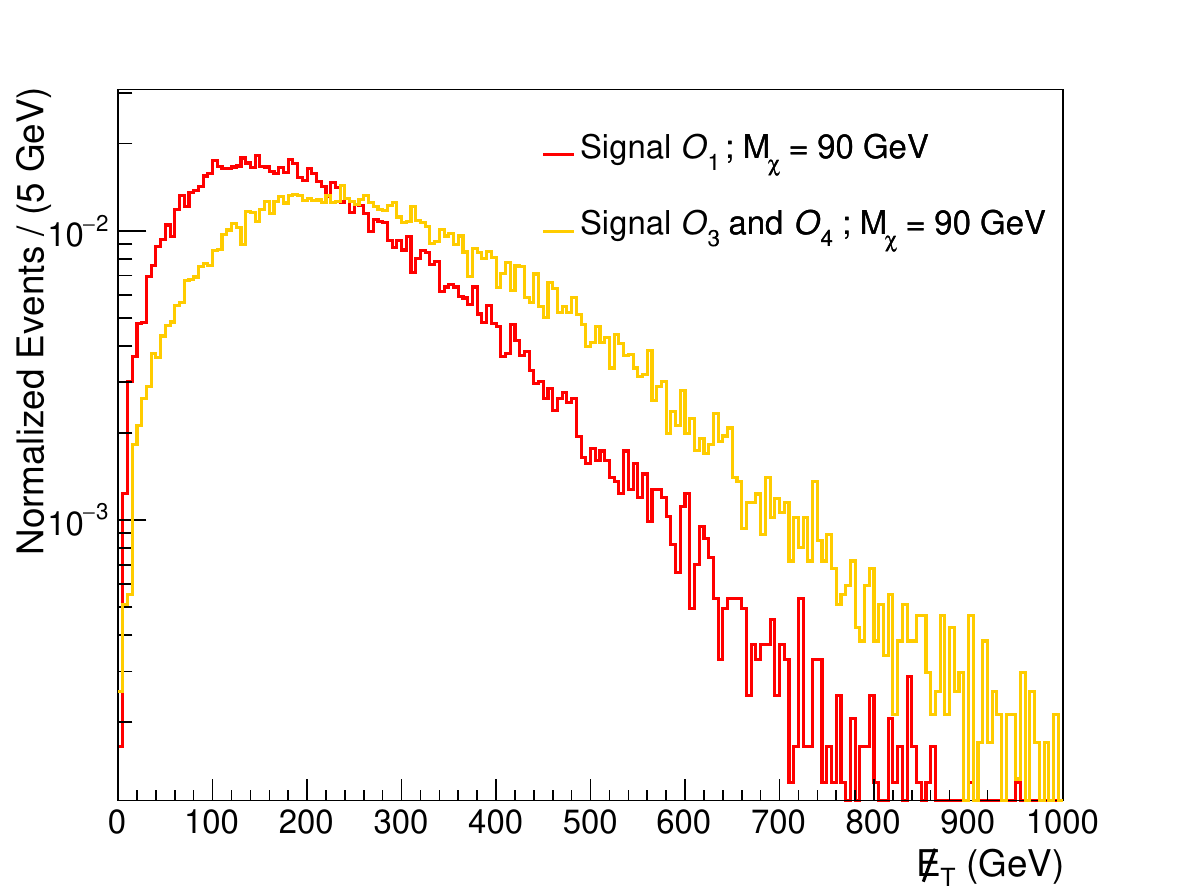}  
        }%
    \end{minipage}
    \begin{minipage}[c]{0.5\linewidth}%
        \vspace{0pt}%
        \centering%
        \subfloat[Invariant mass of the two highest $p_{T}$ b-jets]{%
         \includegraphics[width=\textwidth]{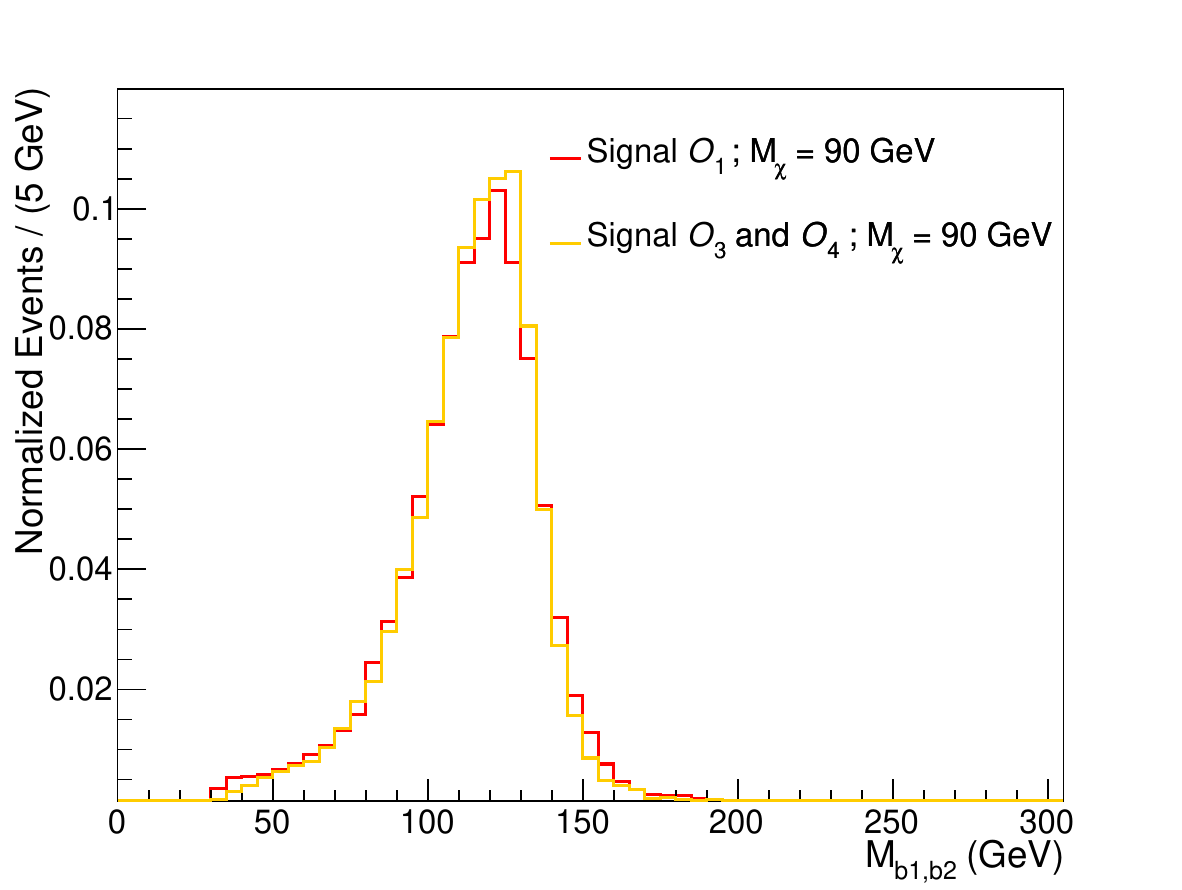}  
        }%
    \end{minipage}
    \caption{Distributions of kinematic variables for signal corresponding to $\mathcal{O}_1$ with $C_1^A/\Lambda^2=15$ TeV$^{-2}$ and combination of $\mathcal{O}_3$ and $\mathcal{O}_4$ with $C_3=C_4$ with $M_{\chi} = 90 ~GeV$, all normalized to unity.}
    \label{fig:Kinematics2}
\end{figure}

We also compare the signal
distributions of $\not\!\! E_{T}$, leading and sub-leading b-jet $p_{T}$, and invariant mass of the b-jet pair for $\mathcal{O}_1$ and combination of $\mathcal{O}_3$ and $\mathcal{O}_4$ with $C_3=C_4$ as shown in Fig.~\ref{fig:Kinematics2} for $M{\chi} = 90 ~GeV$. All the distributions have been normalized to unity. The $p_T$ and $\not\!\! E_{T}$ distributions for combined ${\cal O}_3$ and ${\cal O}_4$ are relatively harder than those for ${\cal O}_1$. This can be explained by the fact that ${\cal O}_3, {\cal O}_4$ are suppressed by another power of $\Lambda$ compared to ${\cal O}_1$. Thus, for the combination of ${\cal O}_3$ and ${\cal O}_4$, cross-section rises more rapidly with energy, resulting in harder $p_T$ and $\not\!\! E_{T}$ distributions.
Guided by the distributions of the key kinematic variables, a cut-based analysis is performed for $\mathcal{O}_1$, which is described in the next section. 

\subsection{Analysis} \label{sec3.2}
The analysis begins with a pre-selection of events having at least two b-tagged jets with medium working point and surviving the lepton veto. We first describe the rectangular cut-based analysis, followed by the one using Boosted Decision Tree\,(BDT)~\cite{Coadou_2022}. For $\mathcal{O}_1$, both cut-based and BDT analysis is performed, while for the combination of $\mathcal{O}_3$ and $\mathcal{O}_4$ we perform only the BDT analysis. The sensitivity of the analysis is demonstrated by computing the signal significance\,(\textbf{S}) for all the BPs, assuming an integrated luminosity\,($\mathcal{L}$) of $3000~fb^{-1}$ at $\sqrt{s}=14$ TeV, expected to be delivered by the HL-LHC. The signal significance is defined as:

\begin{equation}
\textit{\textbf{S}} = \sqrt{2\left[\,(S+B)\ln\left(1+\frac{S}{B}\right)-S\right]}
\label{eq:sigFormula}
\end{equation}
where, $S$ and $B$ are the expected number of signal and background events respectively.

\subsubsection{Cut-Based Analysis} \label{sec3.2.1}
The cut-based analysis has been performed for $\mathcal{O}_1$ by applying selection conditions on the kinematic variables described in Table~\ref{CutBased_variables}, for the events passing the pre-selection mentioned earlier. Table~\ref{KinematicEfficiency} summarizes the cut values along with their efficiencies for signal $\mathcal{O}_1$ with $C_1^A/\Lambda^2=15$ TeV$^{-2}$ and background.
\begin{table}[H]
    \centering
    \renewcommand{\arraystretch}{1.5}
    \begin{tabular}{|c|c|}
	\hline 
    $Variable$ & $Description$ \\ \hline \hline
   $p_{T,b1}$ & $\text{Transverse momentum of the leading b-jet\,($b$1)}$   \\  \hline
   $\Delta{R}(b1,b2)$ & $\text{$\Delta{R}$ between $b$1 and $b$2 }$   \\  \hline
   $\not\!\! E_{T}$ & $\text{Missing Transverse Energy}$  \\ \hline  
   $M_{b1,b2}$ & $\text{Invariant mass of the b-jet pair}$   \\  \hline  
   $p_{T,l1}$ & $\text{Transverse momentum of the leading non b-jet\,($l$1)}$   \\  \hline
   $\Delta{\phi}(b1,\not\!\! E_{T})$ & $\text{$\Delta{\phi}$ between the $b$1 and $\not\!\! E_{T}$}$  \\   \hline
   $\Delta{\phi}(b2,\not\!\! E_{T})$ & $\text{$\Delta{\phi}$ between $b$2 and $\not\!\! E_{T}$}$   \\ \hline
    \end{tabular}
    \caption{List of kinematic variables used in the Cut-Based analysis.}
    \label{CutBased_variables}
\end{table}
\begin{table}[H]
    \centering
    \renewcommand{\arraystretch}{1.5}
    \resizebox{\textwidth}{!}{
    \begin{tabular}{|c|c|c|c|c|c|c|c|c|c|c|c|}
    \hline
    $Process$ & $BP1B$  & $BP2B$ & $BP3B$ & $BP4B$ &  $BP5B$ & $BP6B$ & $ttH$ & $ttL$ &  $ttS$ & $ZZ$ & $Zh$ \\
        \hline 
   \diagbox{Cuts}{Initial Events}
   & \textit{$10^{5}$} & \textit{$10^{5}$} & \textit{$10^{5}$} & \textit{$10^{5}$} & \textit{$10^{5}$} & \textit{$10^{5}$} & \textit{$4\times10^{5}$} & \textit{$3\times10^{5}$} & \textit{$3\times10^{5}$} & \textit{$2\times10^{5}$} & \textit{$10^{5}$}

   \\
   
\hline \hline
$p_{T,b1} > 70~GeV$                       & $26.2$  & $26.5$ & $27.6$ & $28.4$ &  $28.7$ & $28.8$ & $29.4$ & $2.8$ &  $7.4$ & $8.0$ & $16.1$  \\  \hline 
$\Delta{R}(b1,b2) < 1.5$                  & $22.9$  & $23.3$ & $24.9$ & $26.1$ &  $26.6$ & $26.9$ & $23.5$ & $2.3$ &  $6.1$ & $3.6$ & $10.7$  \\ \hline 
$\not\!\! E_{T} > 190~GeV$                & $15.8$  & $16.5$ & $18.8$ & $20.5$ &  $21.5$ & $22.0$ & $4.2$ & $0.3$ &  $1.05$ & $2.1$ & $3.7$  \\  \hline 
$70~GeV < M_{b1,b2} < 150~GeV$            & $11.3$  & $11.9$ & $14.1$ & $15.9$ &  $17.1$ & $17.9$ & \textit{$7.0\times10^{-2}$} & \textit{$4.0\times10^{-2}$} &  $0.1$ & $0.6$ & $1.8$  \\  
\hline 
$p_{T,l1} < 60~GeV$                       & $10.4$  & $10.9$ & $13.0$ & $14.7$ &  $15.7$ & $16.4$ & \textit{$4.0\times10^{-2}$} & \textit{$2.0\times10^{-2}$} &\textit{$6.0\times10^{-2}$} & $0.5$ & $1.6$  \\    
\hline 
$\Delta{\phi}(b1,\not\!\! E_{T}) > 1.0$   & $7.3$   & $7.6$ & $8.9$ & $9.9$ &  $10.5$ & $10.9$ & \textit{$1.5\times10^{-3}$} & \textit{$8.0\times10^{-3}$} & \textit{$9.0\times10^{-3}$}  & $0.4$ & $1.2$  \\    
\hline 
$\Delta{\phi}(b2,\not\!\! E_{T}) > 1.0$   & $7.3$   & $7.6$ & $8.9$ & $9.9$ &  $10.5$ & $10.9$ & $\textit{$1.5\times10^{-3}$}$ & \textit{$8.0\times10^{-3}$} & \textit{$9.0\times10^{-3}$}  & $0.4$ & $1.2$  \\ 
\hline  
  \end{tabular}}
  \caption{Selection cut efficiencies for signal $\mathcal{O}_1$ with $C_1^A/\Lambda^2=15$ TeV$^{-2}$ and backgrounds.}
  \label{KinematicEfficiency}
\end{table}
The signal significance is estimated using Eqn.~\ref{eq:sigFormula}, after all the selection cuts are applied. The obtained values of significance at $\sqrt{s}=14$ TeV assuming $\mathcal{L} = 3000~fb^{-1}$ for all the signal BPs of $\mathcal{O}_1$ are listed in Table~\ref{Significance1}.

\begin{table}[H]
    \centering
    \renewcommand{\arraystretch}{1.5}
    \begin{tabular}{|c|c|c|c|c|c|c|}
    \hline
        $BP~(Signal~\mathcal{O}_1)$ & $BP1B$ & $BP2B$ & $BP3B$ & $BP4B$ & $BP5B$ & $BP6B$ \\ 
        \hline
         $M_{\chi}\,(GeV)$ & $90$ & $100$ & $150$ & $200$ & $250$ & $300$  \\
        \hline \hline
        $\textit{\textbf{S}}$ & $1.7\sigma$ & $1.6\sigma$ & $1.0\sigma$ & $0.7\sigma$ & $0.5\sigma$ & $0.3\sigma$ \\ 
        \hline 
    \end{tabular}
    \caption{Signal significance\,(\textit{\textbf{S}}) for cut-based analysis at $\sqrt{s}=14$ TeV assuming $\mathcal{L} = 3000~fb^{-1}$ for all the $\mathcal{O}_1$ signal benchmark points with $C_1^A/\Lambda^2=15$ TeV$^{-2}$.}
    \label{Significance1}
\end{table}
In order to exploit the signal region further, we used BDT for both $\mathcal{O}_1$ and combination of $\mathcal{O}_3$ and $\mathcal{O}_4$, as described in the next section.

\subsubsection{Analysis using Boosted Decision Tree} \label{sec3.2.2}
\begin{figure}[!htp]
    \begin{minipage}[c]{0.33\linewidth}%
        \vspace{0pt}%
        \centering%
        \subfloat[Number of jets]{%
         \includegraphics[width=\textwidth]{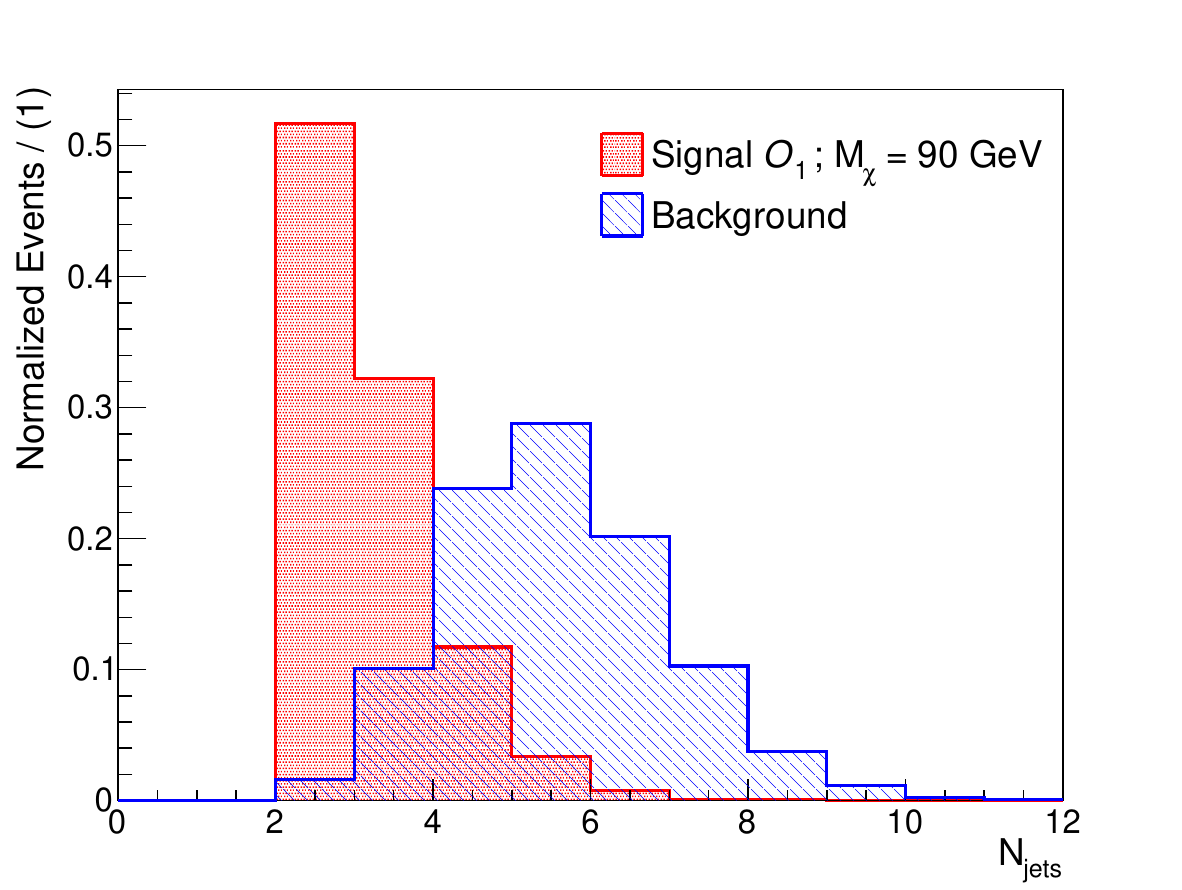}
        }%
    \end{minipage}
    \begin{minipage}[c]{0.33\linewidth}%
        \vspace{0pt}%
        \centering%
        \subfloat[$p_{T}$ of leading b-jet]{%
         \includegraphics[width=\textwidth]{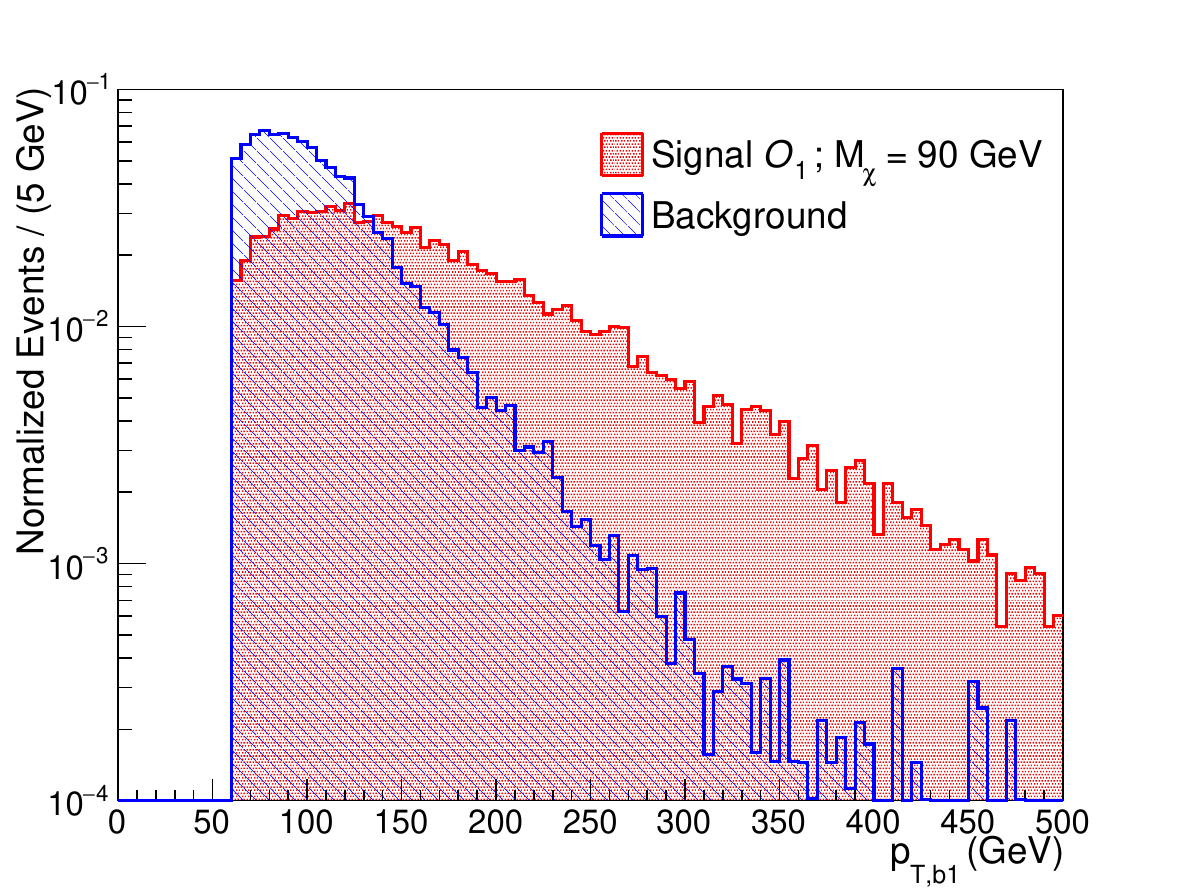}
        }%
    \end{minipage}   
    \begin{minipage}[c]{0.33\linewidth}%
        \vspace{0pt}%
        \centering%
        \subfloat[$\Delta{R}$ between $b$1 and $b$2 ]{%
        \includegraphics[width=\textwidth]{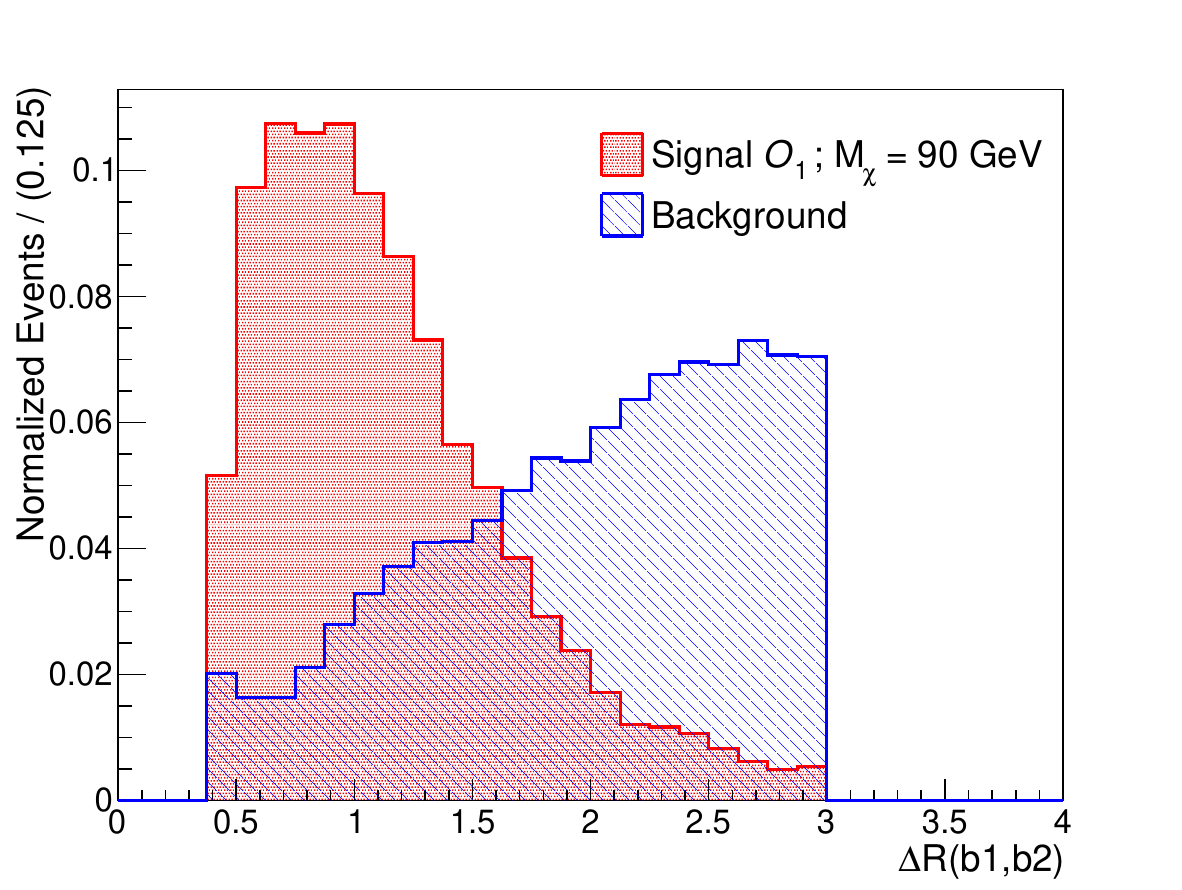}
        }%
    \end{minipage}  
    \begin{minipage}[c]{0.33\linewidth}%
        \vspace{0pt}%
        \centering%
        \subfloat[Invariant mass of $b$1 and $b$2 ]{%
        \includegraphics[width=\textwidth]{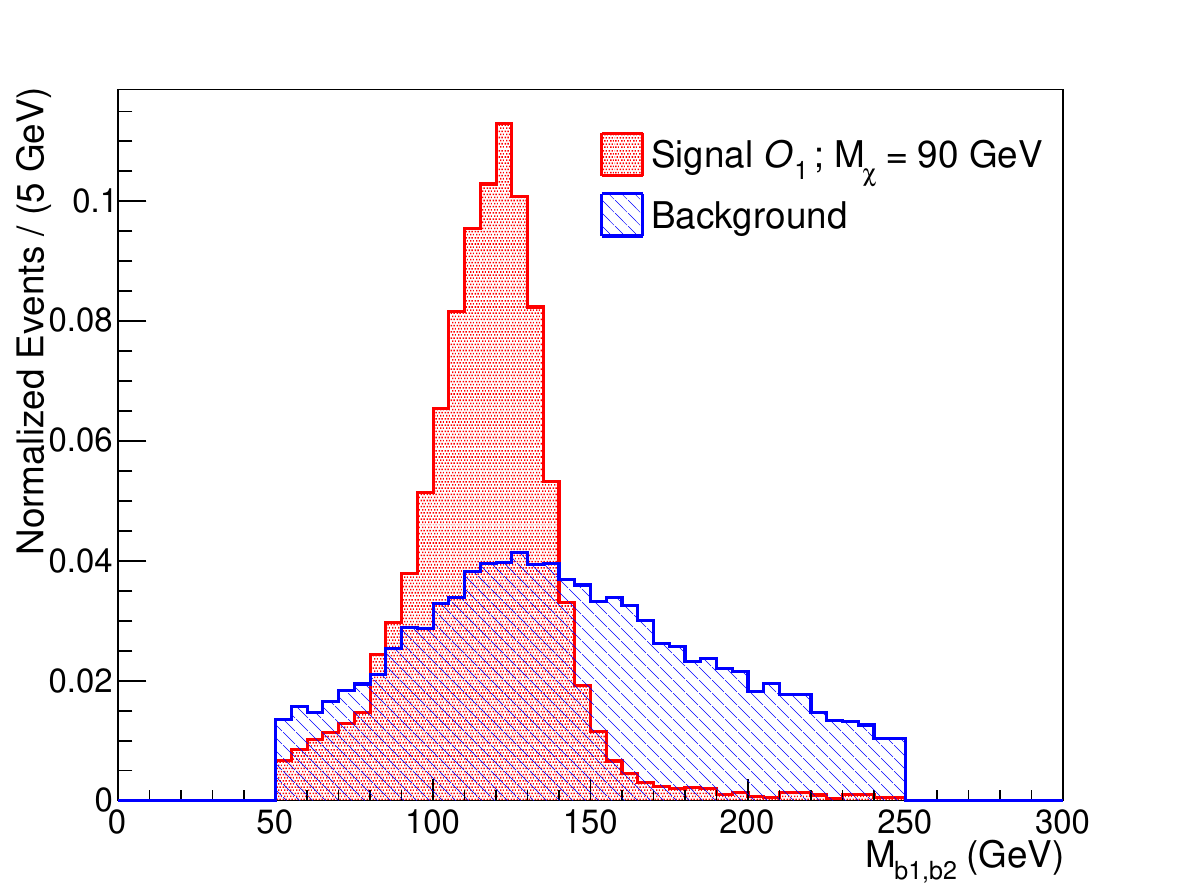}
        }%
    \end{minipage}    
    \begin{minipage}[c]{0.33\linewidth}%
        \vspace{0pt}%
        \centering%
        \subfloat[$\Delta{\phi}$ between $b$1 and $l$1]{%
         \includegraphics[width=\textwidth]{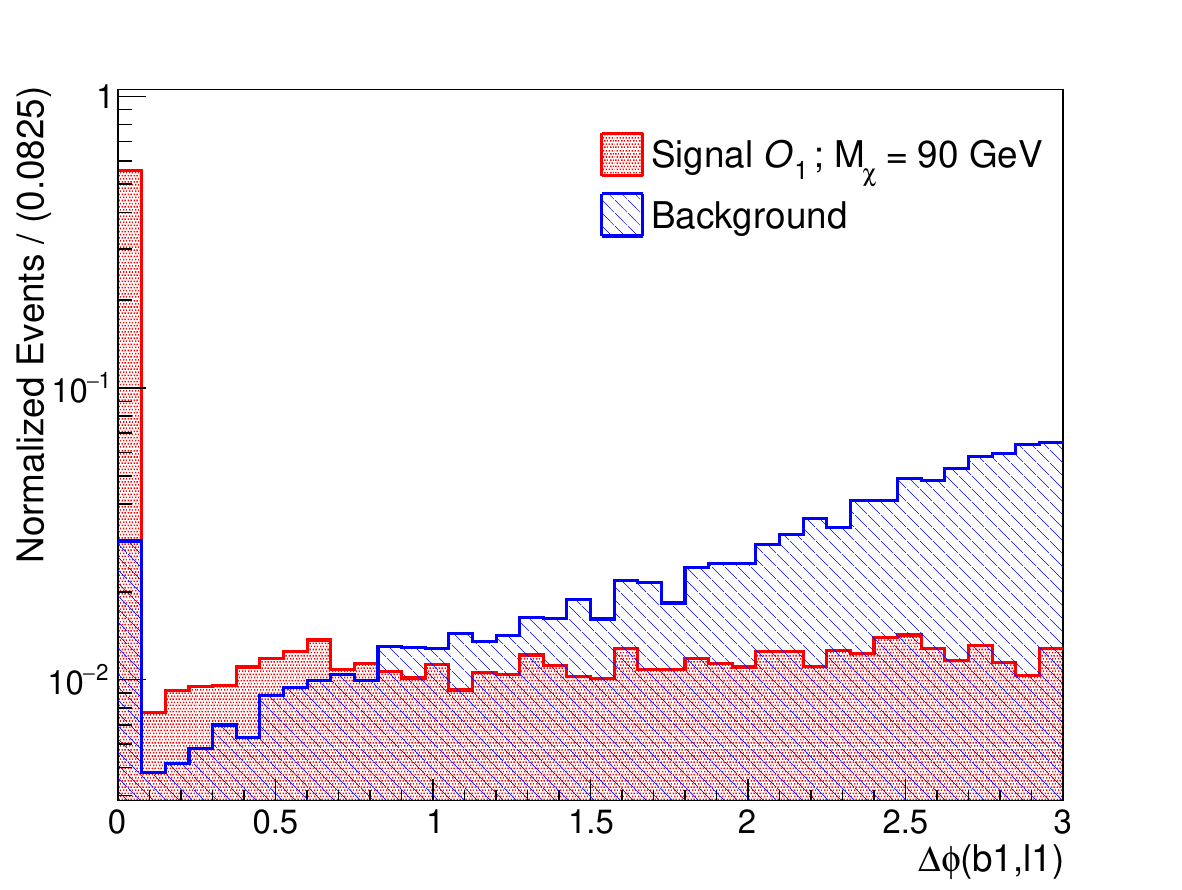}
        }%
    \end{minipage}     
    \begin{minipage}[c]{0.33\linewidth}%
        \vspace{0pt}%
        \centering%
        \subfloat[$\Delta{\phi}$ between $b$1 and $\not\!\! E_{T}$]{%
         \includegraphics[width=\textwidth]{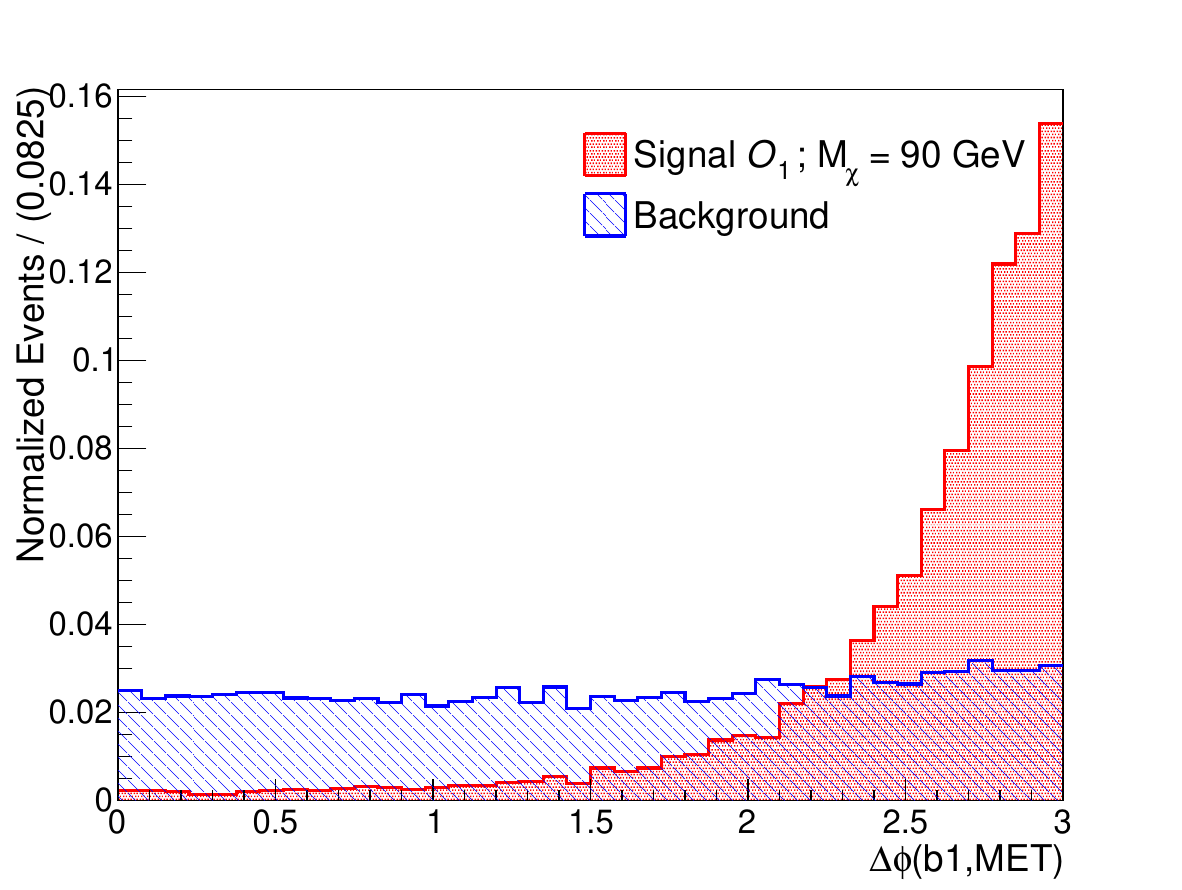}
        }%
    \end{minipage}
    \begin{minipage}[c]{0.33\linewidth}%
        \vspace{0pt}%
        \centering%
        \subfloat[$\Delta{\phi}$ between $b$2 and $\not\!\! E_{T}$]{%
         \includegraphics[width=\textwidth]{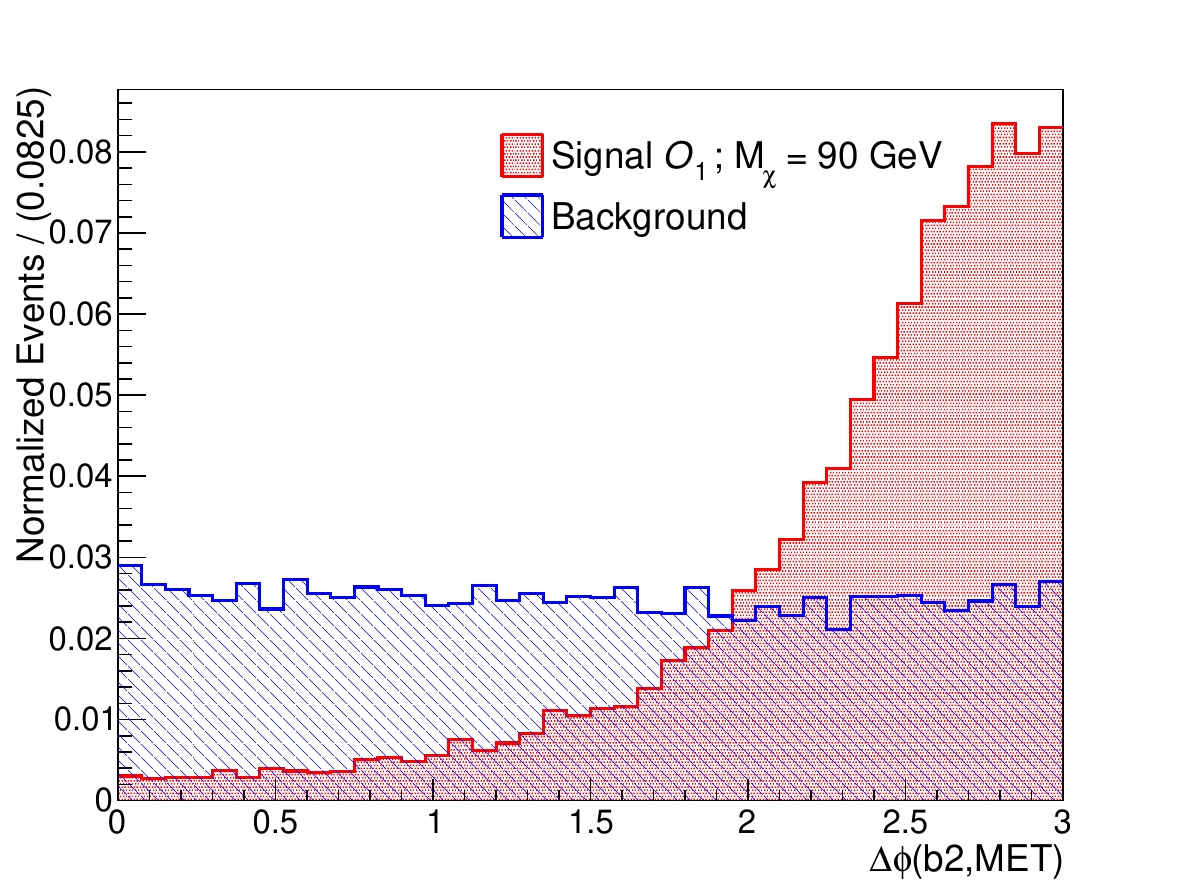}
        }%
    \end{minipage}     
    \begin{minipage}[c]{0.33\linewidth}%
        \vspace{0pt}%
        \centering%
        \subfloat[Vector Sum of $p_{T}$ of all jets]{%
         \includegraphics[width=\textwidth]{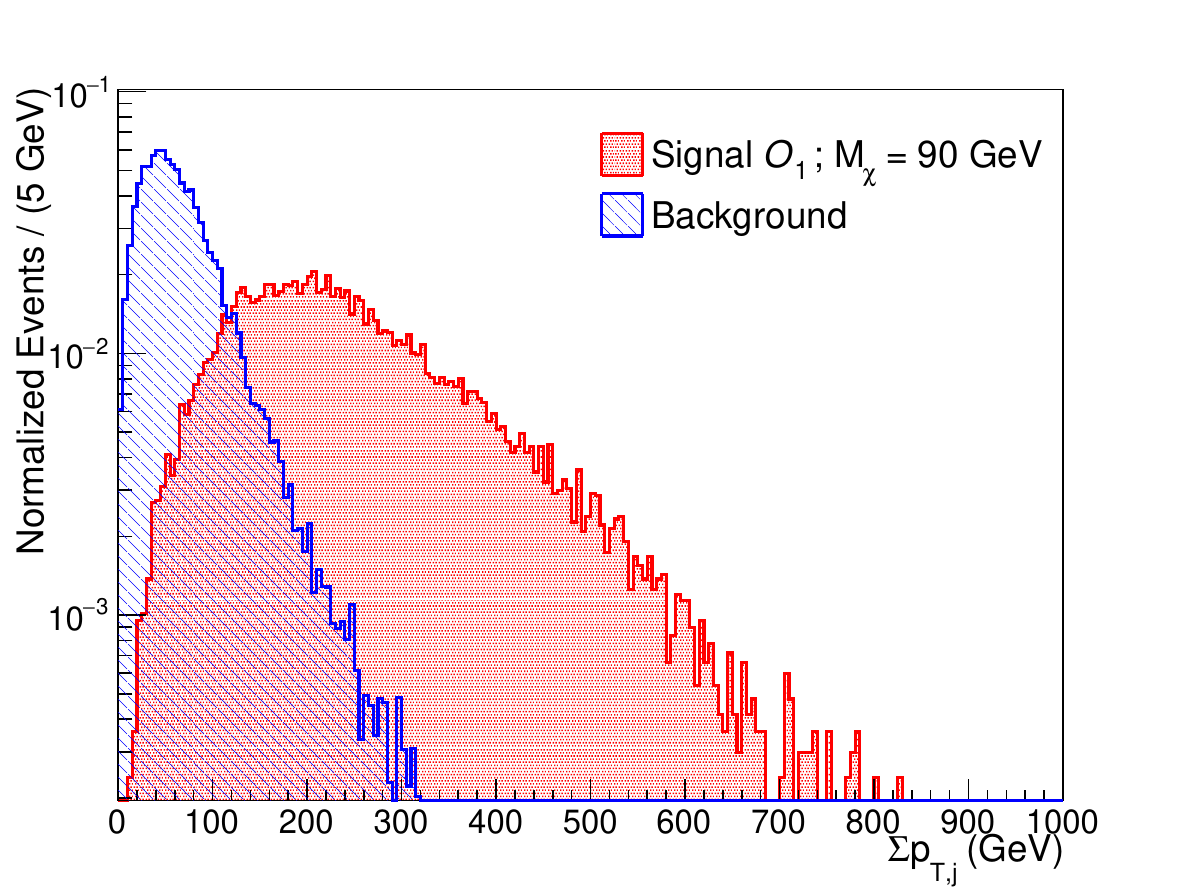}
        }%
    \end{minipage}     
    \begin{minipage}[c]{0.33\linewidth}%
        \vspace{0pt}%
        \centering%
        \subfloat[$\not\!\! E_{T}$ significance]{%
         \includegraphics[width=\textwidth]{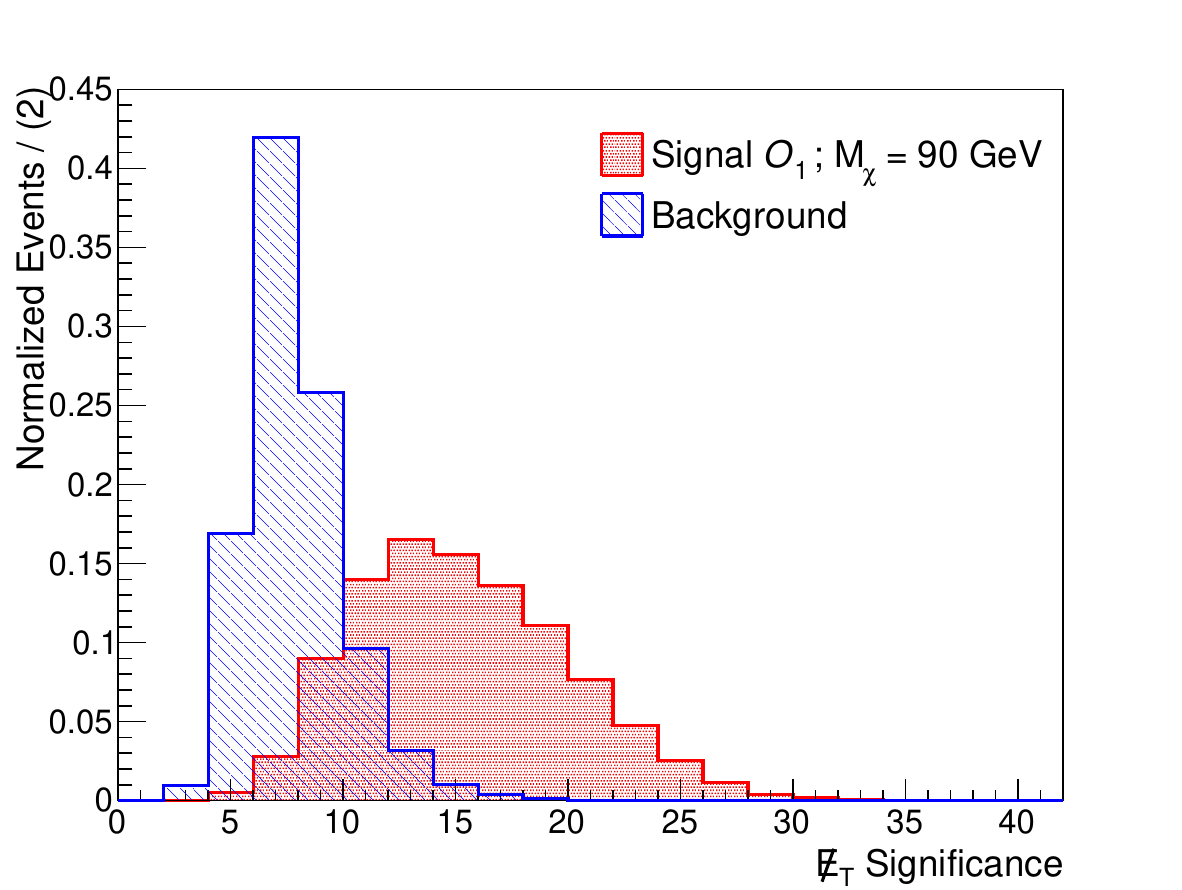}
        }%
    \end{minipage}   
    \caption{Distribution of feature variables taken as input for BDT in Signal $\mathcal{O}_1$ BP1B and background.}
    \label{fig:BDTInput}
\end{figure}

 The BDT algorithm is a supervised multivariate technique used in event classification problems. It optimizes the separation of signal and background through an ensemble of decision trees. We have used the TMVA-Toolkit for Multivariate Data Analysis~\cite{hoecker2009tmva} package for our study. All the kinematic variables used in the cut-based analysis\,(Table \ref{CutBased_variables}), except $\not\!\! E_{T}$ and $p_{T,l1}$, a number of new variables defined in Table~\ref{BDT_variables}, are used in the BDT.
\begin{table}[H]
    \centering
    \renewcommand{\arraystretch}{1.5}
    \begin{tabular}{|c|c|}
	\hline 
        $Variable$ & $Description$ \\ 
	\hline \hline
	$N_{jets}$ & $\text{Total number of jets}$   \\ 
	\hline            
        $p_{T,b2}$ & $\text{Transverse momentum of the sub-leading b-jet\,($b$2)}$   \\             
        \hline  
 	$\Delta{\phi}(b1,l1)$ & $\text{$\Delta{\phi}$ between $b$1 and leading non b-jet\,($l$1)}$   \\ 
        \hline      
 	$\Sigma p_{T,lj}$ & $\text{Scalar sum of $p_{T}$ of light jets}$   \\ 
        \hline
 	$\Sigma \vv p_{T,j}$ & $\text{Vector sum of $p_{T}$ of all jets}$   \\ 
        \hline   
        $\not\!\! E_{T}/\sqrt{\vv p_{T,j}}$ & $\text{$\not\!\! E_{T}$ significance defined as $\not\!\! E_{T}/\sqrt{\vv p_{T,j}}$}$   \\ 
        \hline   
    \end{tabular}
    \caption{List of kinematic variables used in the BDT based analysis, in addition to the ones mentioned in the cut-based analysis.}
    \label{BDT_variables}
\end{table}
 The following additional loose selection cuts have been applied on the pre-selected events before passing them through BDT:
\begin{itemize}
    \item $p_{T,b1}> 60~GeV$
    \item $\Delta{R}(b1,b2) < 3$ 
    \item $\not\!\! E_{T} > 120~GeV$
    \item $50~GeV < M_{b1,b2} < 250~GeV$
\end{itemize}
Fig.~\ref{fig:BDTInput} shows the normalized distribution of the input variables to BDT for signal from $\mathcal{O}_1$ BP1B\,(blue-dashed) and all background\,(red-shaded) processes.

\clearpage

Table~\ref{BDT_parameters} shows the values of the BDT parameters used in this analysis. The network is trained and tested separately for all the benchmark points for both the signals. Sixty percent events are used for training the BDT for both signal and background. The background events are weighted by the cross-sections of the corresponding processes and combined during training. The variables are ranked based on how frequently they are used to split nodes in the decision trees. $\Sigma \vv p_{T,j}$ and $\not\!\! E_{T}~significance$ are found to be the two most significant variables for signal and background discrimination. The next most important variable is $\Delta{\phi}(b2,\not\!\! E_{T})$ for $\mathcal{O}1$ and $M_{b1,b2}$ for the combination of $\mathcal{O}_3$ and $\mathcal{O}_4$.
\begin{table}[H]
    \centering
    \renewcommand{\arraystretch}{1.5}
    \begin{tabular}{|c|c|c|}
	\hline 
BDT parameters & Description & Value\\
\hline
\texttt{NTrees} & Number of trees or nodes & $850$ \\
\texttt{MinNodeSize} & Minimum $\%$ of training events required in a leaf node & $2.5 \%$\\
\texttt{MaxDepth} & Max depth of the decision tree allowed & $3$\\
\texttt{BoostType} & Boosting mechanism to make the classifier robust & Gradient Boost \\
\texttt{Shrinkage} & Learning rate for Gradient algorithm & $0.5$\\
\texttt{nCuts} & Number of grid points in variable range & \\
                              & used in finding optimal cut in node splitting & $30$ \\
\hline
\end{tabular}
\caption{List of BDT parameters and their used values.}
\label{BDT_parameters}
\end{table}

\begin{figure}[!htp]
    \begin{minipage}[c]{0.5\linewidth}%
        \vspace{0pt}%
        \centering%
        \subfloat[Signal $\mathcal{O}_1$]{%
        \includegraphics[width=\textwidth]{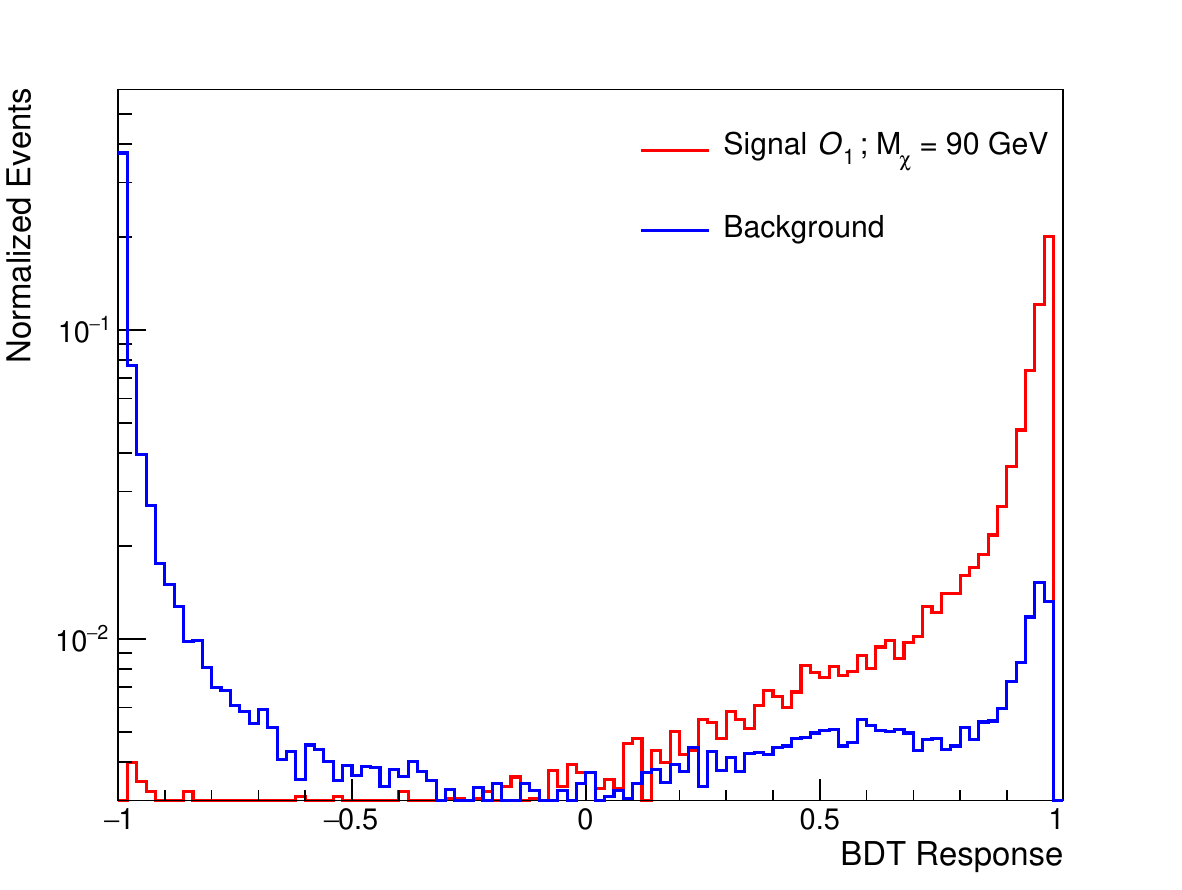}
        }%
    \end{minipage}
    \begin{minipage}[c]{0.5\linewidth}%
        \vspace{0pt}%
        \centering%
        \subfloat[Signal $\mathcal{O}_3$ and $\mathcal{O}_4$]{%
         \includegraphics[width=\textwidth]{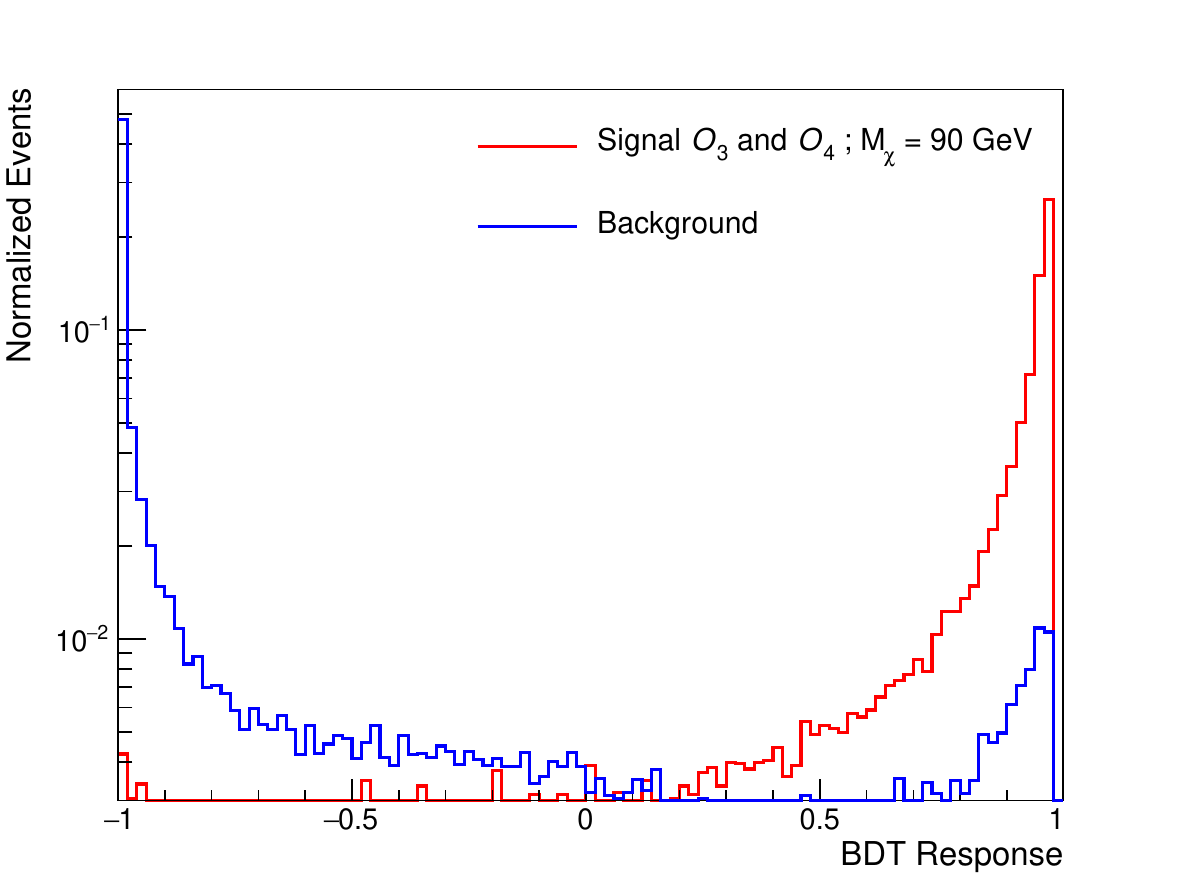}
         }%
    \end{minipage}
    \caption{Distribution of BDT response for BP1B\,(left) and BP1C\,(right) signal and background, all normalized to unity.}
    \label{fig:BDTResponse}
\end{figure}

\begin{figure}[!htp]
    \begin{minipage}[c]{0.5\linewidth}%
        \vspace{0pt}%
        \centering%
        \subfloat[Signal $\mathcal{O}_1$]{%
        \includegraphics[width=\textwidth]{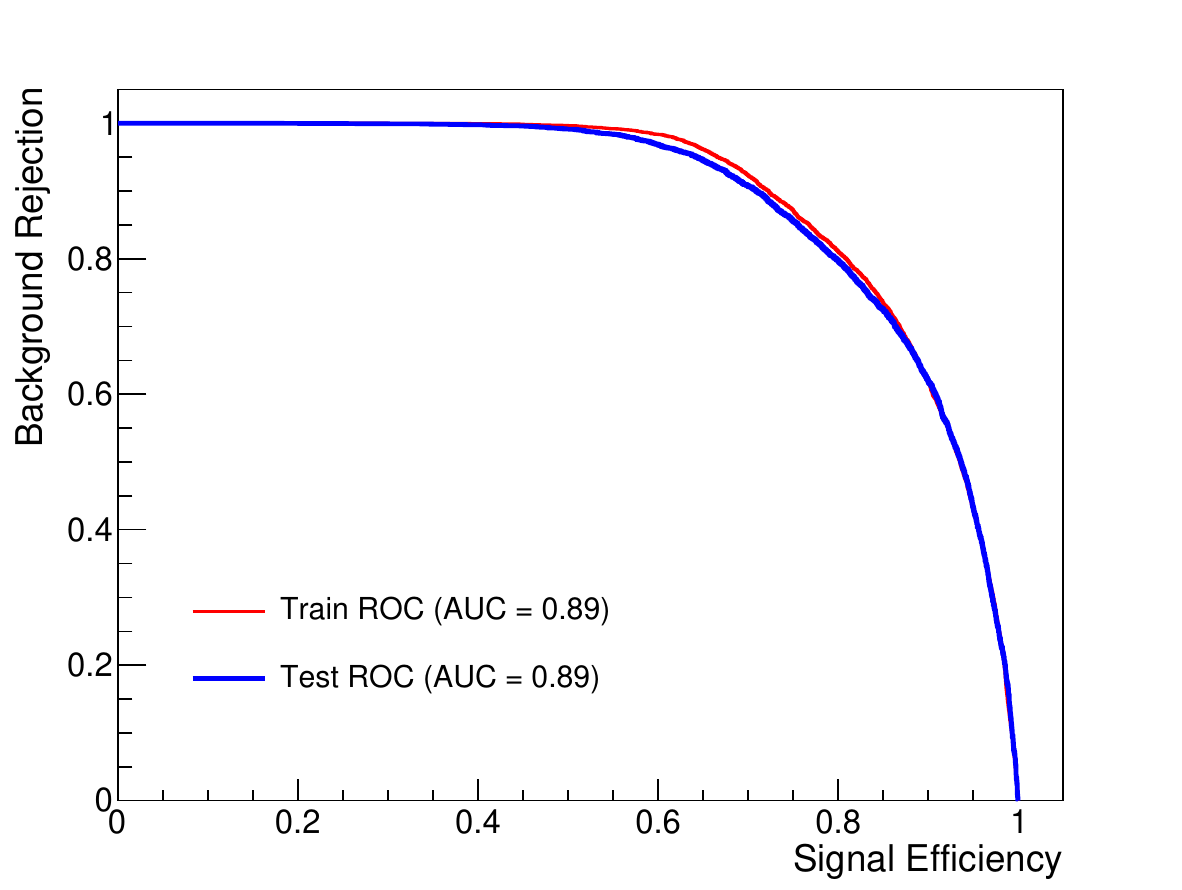}
        }%
    \end{minipage}
    \begin{minipage}[c]{0.5\linewidth}%
        \vspace{0pt}%
        \centering%
        \subfloat[Signal $\mathcal{O}_3$ and $\mathcal{O}_4$]{%
         \includegraphics[width=\textwidth]{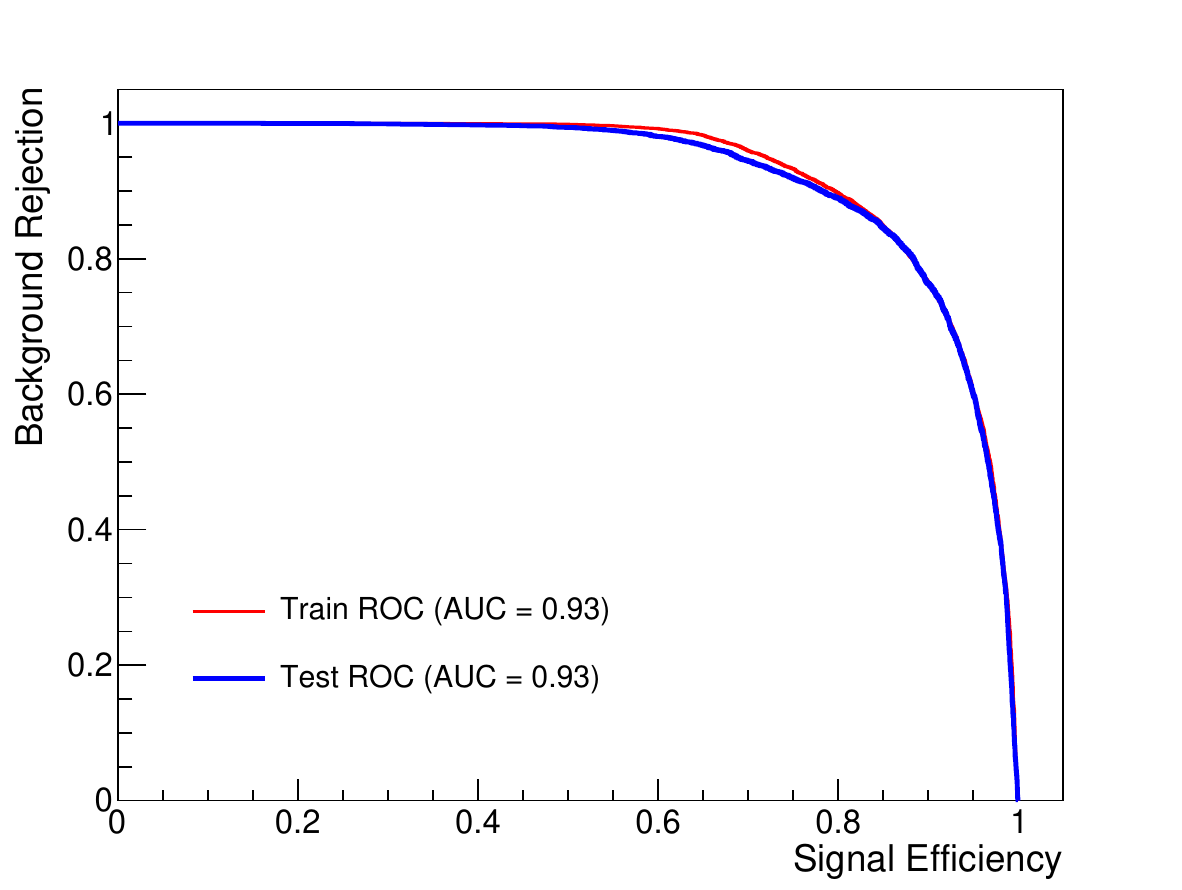}
         }%
    \end{minipage}
    \caption{ROC curves for $\mathcal{O}_1$ BP1B signal\,(left) and combination of $\mathcal{O}_3$ and $\mathcal{O}_4$ BP1C signal\,(right) with their AUC values.}
    \label{fig:BDTROC}
\end{figure}

The algorithm yields a well discriminating profile of the BDT response for the signal and backgrounds processes. The distribution of the final BDT response comparing both the signals\,($\mathcal{O}_1$ and combined $\mathcal{O}_3$ and $\mathcal{O}_4$), with total background can be seen in Fig.~\ref{fig:BDTResponse}. Fig.~\ref{fig:BDTROC} shows the Receiver Operating Characteristic\,(ROC) curve for the BDT.
The area under the curve\,(AUC) values for  training\,(test) data samples are 0.89\,(0.88) and 0.93\,(0.93) for $\mathcal{O}_1$ and combined $\mathcal{O}_3$ and $\mathcal{O}_4$ signals respectively. The ROC curves for both the training and test samples are quite similar with closely matching AUC values, indicating minimal over-training. The BDT response is then utilized to estimate the signal significance using Eqn.\,\ref{eq:sigFormula}. The obtained values are listed in Table\,\ref{Significance2}.
\begin{table}[H]
    \centering
    \renewcommand{\arraystretch}{1.5}
    \begin{tabular}{|c|c|c|c|c|c|}
	\hline 
    & $M_{\chi}$\,(GeV) & \makecell{\textbf{A} \\ \textit{Signal $\mathcal{O}_1$} \\ \textit{$\dfrac{C_{1}^A}{\Lambda^2} = 8.75$} TeV$^{-2}$} & \makecell{\textbf{B} \\ \textit{Signal $\mathcal{O}_1$} \\ \textit{$\dfrac{C_{1}^A}{\Lambda^2} = 15$} TeV$^{-2}$} & \makecell{\textbf{C} \\ \textit{Signal $\mathcal{O}_3~and~\mathcal{O}_4$} \\ \textit{$\dfrac{C_3}{\Lambda^3}=\dfrac{C_4}{\Lambda^3} = 5$} TeV$^{-3}$} & \makecell{\textbf{D} \\ \textit{Signal $\mathcal{O}_3~and ~\mathcal{O}_4$} \\ \textit{$\dfrac{C_3}{\Lambda^3}=\dfrac{1}{2}\dfrac{C_4}{\Lambda^3}= 2$} TeV$^{-3} $}  \\ 
	\hline \hline
	BP1 & $90$ & $0.89\sigma$ & $2.6\sigma$ & $5.5\sigma$ & $2.76\sigma$ \\ 
    \hline
 	BP2 & $100$ & $0.85\sigma$ & $2.5\sigma$ & $4.9\sigma$ & $2.42\sigma$ \\ 
    \hline
    BP3 & $150$ & $0.58\sigma$ & $1.7\sigma$ & $4.7\sigma$ & $2.35\sigma$ \\ 
    \hline
 	BP4 & $200$ & $0.41\sigma$ & $1.2\sigma$ & $4.3\sigma$ & $2.22\sigma$ \\ 
    \hline
 	BP5 & $250$ & $0.24\sigma$ & $0.7\sigma$ & $3.5\sigma$ & $1.79\sigma$ \\
    \hline
    BP6 & $300$ & $0.17\sigma$ & $0.5\sigma$ & $3.3\sigma$ & $1.70\sigma$ \\ 
        \hline 
    \end{tabular}
    \caption{Signal significance for BDT at $\sqrt{s}=14$ TeV assuming $\mathcal{L} = 3000~fb^{-1}$ for all the signal benchmark points.}
     \label{Significance2}
\end{table}

The signal significance improves for all the BPs\,(set \textbf{B}) with BDT for  $\mathcal{O}_1$ compared to the cut-based analysis which can be seen in Tables~\ref{Significance1} and~\ref{Significance2}.

\section{Summary and Conclusion} \label{sec4}
To summarize, we have performed a model independent study to investigate the prospects of discovering the relic particle at the high luminosity run of the LHC. The proposed relic particle, a Dirac fermion in our framework, has no charge under the SM gauge group. However,
its mass is well within the kinematic reach of the HL-LHC. Such a particle can arise in an UV complete BSM theory alongwith a plethora of other particles, which are possibly too heavy and are beyond the reach of the LHC. 
An appropriate way to describe the physics of such a scenario is to use the framework of EFT, where the fields with masses comparable to the energy scale of interest are the only relevant degrees of freedom and the fields which are much heavier are integrated out. Consequently, one is left with certain higher\,$(> 4)$ dimensional operators which are suppressed by appropriate powers of energy scale and have unknown coefficients. We have followed the same prescription by constructing dim-6 and dim-7 operators involving the relic particle, $\chi$, the Higgs boson, and the gauge bosons of the SM.
These operators could be viewed as the only portal to the dark sector and are responsible for any interaction of the relic particle with the SM particles. 
Consequently, the unknown Wilson coefficients of the proposed operators could be constrained from the measurement of relic density and experimental upper limits on the DM-nucleon scattering cross-section.

As a first step, we determined the allowed range of values of the Wilson coefficients by comparing the model predictions for the aforementioned quantities with the experimental data scanned over a range of DM mass starting from 10 GeV to 800 GeV. Results from XENONnT excludes DM masses below 66 GeV for dim-7 operator for combination of $\mathcal{O}_3$ and $\mathcal{O}_4$ with $C_3$ = $C_4$. Next, we set to explore the prospects of producing the relic particles in association with the SM Higgs boson via higher dimensional interactions at the LHC. The Higgs boson subsequently decays to a pair of $b$-quarks resulting in two $b$-jets plus $\not\!\! E_{T}$ in the final state. Similar final state can arise from various SM processes like $t \bar t$, $ZZ$ and $Zh$. 
Using simulation samples for signal and SM backgrounds, we have performed both cut-based and BDT based analysis in the HL-LHC scenario. Considering a projected integrated luminosity of $\mathcal{L} = 3000~fb^{-1}$ to be recorded at the HL-LHC by the CMS experiment, our results show that for the combination of signals $\mathcal{O}_3$ and $\mathcal{O}_4$ with $C_3/\Lambda^3 = C_4/\Lambda^3 = 5$ TeV$^{-3}$\,(set \textbf{C}), significances of 5.5$\sigma$ and 4.9$\sigma$ are obtained for $M\chi = 90$ GeV\,(BP1) and $100$ GeV\,(BP2), respectively. For $M_\chi$ values in the range of 90-300 GeV studied in this article, the signal significance varies from 5.5$\sigma$ to 3.3$\sigma$. For the combination of signals $\mathcal{O}_3$ and $\mathcal{O}_4$ with $C_3/\Lambda^3 = \dfrac{1}{2}C_4/\Lambda^3 = 2$ TeV$^{-3}$\,(set \textbf{D}), the significance ranges from 2.76$\sigma$ to 1.7$\sigma$. For signal $\mathcal{O}_1$ with $C_1^A/\Lambda^2 = 15$ TeV$^{-2}$ (set \textbf{B}), the significance varies between 2.6$\sigma$ and 0.5$\sigma$, while for set \textbf{A}, it remains below 1$\sigma$. This work highlights the potential of a model independent search for DM at the HL-LHC and serves as a basis for future investigations in this area.

\section*{Acknowledgement}
We acknowledge the central computing facility of Saha Institute of Nuclear Physics for computational support.
The work of SBh is supported by the Core Research Grants CRG/2021/007579 and CRG/2022/001922 of the Science and Engineering Research Board (SERB), Government of India.

	\providecommand{\href}[2]{#2}\begingroup\raggedright\endgroup
	

\end{document}